\documentclass{emulateapj}

\shorttitle{SN Delay Times in Galaxy Clusters}
\shortauthors{Maoz et al.}
\begin{document}

\title{The Supernova Delay Time Distribution in Galaxy Clusters and
  Implications for Type-Ia Progenitors and Metal Enrichment
\altaffilmark{*}}
\altaffiltext{*}{Based on observations made with the NASA/ESA 
  {\it Hubble Space Telescope}, obtained at the Space Telescope Science
  Institute, which is operated by the Association of Universities for
  Research in Astronomy, Inc., under NASA contract NAS 5-26555. These
  observations are associated with programs GO-10493 and GO-10793.} 
\author{{Dan Maoz\altaffilmark{1}},
Keren Sharon\altaffilmark{2}, 
{Avishay Gal-Yam\altaffilmark{3}}
}

\altaffiltext{1}{School of Physics and Astronomy, Tel Aviv University,
  Tel Aviv 69978, Israel.} 
\altaffiltext{2}{Kavli Institute for Cosmological Physics, The
  University of Chicago, Chicago, IL 60637.} 
\altaffiltext{3}{Benoziyo Center for Astrophysics, Faculty of Physics,
  Weizmann Institute of Science, Rehovot 76100, Israel.}

\begin{abstract}
Knowledge of the supernova (SN) delay time distribution (DTD) -- the SN
rate versus time that would follow a hypothetical 
brief burst of star formation --
can shed light on SN progenitors and physics, as well as on the timescales
of chemical enrichment in different environments.
 We compile
recent measurements of the Type-Ia SN (SN~Ia) rate in galaxy clusters at
redshifts from $z=0$ out to $z=1.45$, just 2~Gyr after cluster star 
formation at $z\approx 3$.  
We review the plausible range for the observed
total iron-to-stellar mass ratio in clusters, based on the latest data and 
analyses, and use it to constrain the time-integrated number 
of SN~Ia events in 
clusters. With these data, we recover the 
DTD of SNe~Ia in cluster environments. 
The DTD is sharply peaked at the shortest
time-delay interval we probe, $0<t<2.2$~Gyr, with a low tail 
out to delays of $\sim 10$~Gyr, and is remarkably consistent with 
several recent DTD reconstructions based on different methods, applied to
different environments.   
We 
test DTD models from the literature, requiring that they
simultaneously reproduce the observed cluster SN rates and the
observed iron-to-stellar mass ratios. A parametrized 
power-law DTD of the form $t^{-1.2\pm0.3}$ from $t=400$~Myr 
to a Hubble time, can satisfy both
  constraints. Shallower power laws, such as $t^{-1/2}$ cannot,
  assuming a single DTD, and a single star-formation burst (either brief or 
extended) at high $z$. This implies 50--85\% of SNe~Ia explode within
1~Gyr of star formation. 
DTDs from double-degenerate (DD) models, which 
 generically have $\sim t^{-1}$ shapes 
over a wide range of timescales, match the data, but only if
 their predictions
are scaled up by factors of $5-10$. Single degenerate (SD) DTDs always
give poor fits to the data, due to a lack of delayed SNe and overall low 
numbers  of SNe.
The observations can also be reproduced with a combination
of two SN~Ia populations --  a prompt SD
population of SNe~Ia that explodes within a few Gyr of star
formation, and produces about 60 percent  of the iron mass in
clusters, and  a DD population
that contributes the events seen at $z<1.5$
An alternative scenario of a single, prompt, SN~Ia population, but a composite
star-formation history in clusters, consisting of a burst at high $z$,
followed by a constant star-formation rate, can reproduce the SN
rates, but is at odds with direct measurements
 of star formation in clusters at $0<z<1$. Our results support the
 existence of a DD progenitor channel for SNe~Ia, if the overall
 predicted numbers can be suitably increased.

\end{abstract}


\section{Introduction}
\label{ss.intro}
Supernovae (SNe) play a central role in astrophysics, not only as
distance indicators for cosmology (e.g., Riess et al. 1998; Perlmutter
et al. 1999), but as prime synthesizers of heavy
elements (e.g. Woosley 2007), 
sources of kinetic energy, and accelerators of cosmic rays
(e.g. Helder et al. 2009). 
However, many of the most basic facts about these events are still
poorly understood. Core-collapse (CC) SNe are descended from massive stars,
roughly in the range $8-50~{\rm M_\odot}$ (e.g. Gal-Yam \& Leonard
2009), but the exact limits are not
known, neither from theory nor from observation. It is not yet clear 
what are the physical parameters (mass, binarity, rotation, magnetic
field, and more) that determine the variety of observed CC-SN subtypes
(IIP, Ib, Ic, IIn, IIL; see Filippenko 1997, for a review). 
Type-Ia SNe (SNe~Ia) are linked by indirect
evidence to the thermonuclear detonations of  carbon-oxygen white
dwarfs (WDs) whose mass has grown to near the Chandrasekhar limit
(Hoyle \& Fowler 1960). However,
competing scenarios exist for the initial conditions and evolutionary 
paths that lead to this mass growth. In the single degenerate (SD)
model (Whelan \& Iben 1974) a WD grows in mass through accretion from
a non-degenerate stellar companion.  In the
double degenerate (DD) model (Iben \& Tutukov 1984; Webbink 1984), 
two WDs merge after losing energy and 
angular momentum to gravitational waves. 
In both scenarios, many question remain  regarding the ignition and 
development of
the explosion itself. It is only recently that 
the relative and absolute rates of CC SNe and
SNe~Ia as a function of environment and cosmic time are starting to be measured
accurately, and therefore the respective quantity, types, and
times of their contributions to 
 metal enrichment history are still poorly constrained.

A fundamental function that could shed light on all of these issues is    
the SN delay time distribution (DTD). The DTD is the hypothetical SN
rate versus time that would follow a brief burst of star formation. 
The DTD is directly linked to the lifetimes (i.e.,
the initial masses) of the progenitors and to the 
binary evolution timescales up to the
explosion, and therefore different progenitor scenarios predict
different DTDs. Furthermore, the DTDs of different SN types dictate
directly the mix of contributions of different SN types to metal
enrichment throughout cosmic history. Until recently, only few, and
often-contradictory, observational constraints on the DTD existed.
One observational 
approach has been to compare the SN rate in field galaxies, as a
function of redshift, to the cosmic star formation history (SFH). Given that
the DTD is the SN ``response'' to a short burst of star formation, 
the SN rate versus cosmic time will be the convolution of the full
SFH with the DTD. Gal-Yam \& Maoz (2004) carried
out the first such comparison, using a small sample of SNe~Ia out to
$z=0.8$, and concluded that the results were strongly dependent on the
poorly known cosmic SFH, 
a conclusion echoed by Forster et al. (2006).

 With the availability of SN rate measurements
to higher redshifts, Barris \& Tonry (2006) found a SN~Ia rate that
closely tracks the SFH out to $z\sim 1$, and concluded  that the DTD
must be concentrated at short delays, $\lesssim 1$~Gyr. Similar
conclusions have been reached, at least out to $z\sim 0.7$, by
Sullivan et al. (2006). In contrast,
Dahlen et al. (2004, 2008) and Strolger (2004) have argued for a DTD
that is peaked at a delay of $\sim 3$~Gyr, with little power at short
delays, based on a decrease in the SN~Ia rate at $z>1$. However,
Kuznetzova et al. (2007) have re-analyzed some of these datasets and
concluded that the small numbers of SNe and their potential
classification  errors preclude reaching a conclusion. Similarly,
Poznanski et al. (2007) performed new measurements of the $z>1$ SN~Ia
rate, and found that, within uncertainties, the SN rate could be
tracking the SFH. This, again, would imply a short delay time.
Greggio et al. (2008) pointed out that underestimated extinction
of the highest-$z$ SNe, observed in their rest-frame ultraviolet
emission, could be an additional factor affecting these results.
  
A second approach to recovering the DTD has been to compare the SN
rates in galaxy populations of different characteristic ages. Using
this approach, Mannucci
et al. (2005, 2006), Scannapieco \& Bildsten (2005), and Sullivan
(2006) all found evidence for the co-existence of two SN~Ia
populations, a ``prompt'' population that explodes within  $\sim
100-500$~Myr, and a delayed channel that produces SNe~Ia on timescales of
order 5~Gyr. Naturally, these two ``channels'' may in reality be just 
integrals over a continuous DTD on two sides of some time border
(Greggio et al. 2008). 
Totani et al. (2008) have used a similar approach to recover the DTD, by
comparing SN~Ia rates in early-type galaxies of different
characteristic ages, seen at $z=0.4-1.2$ as
part of the Subaru/XMM-Newton Deep Survey (SXDS) project.
They find a DTD consistent with a $t^{-1}$ form, 
which is roughly generic for DD models (e.g. Greggio 2005; see 
\S\ref{sspowerlawdtd}, below). Additional 
recent attempts to address the issue with the ``rate vs. age''
approach have been made by Aubourg et al. (2008), Raskin et al. 
(2009), Yasuda \& Fukugita
(2009), and Cooper et al. (2009).

Both of the above approaches involve averaging, and hence some loss
of information. In the first approach, one averages over large galaxy
populations, by associating all of the SNe detected at a given
redshift with all the galaxies of a particular type at that redshift.
In the second approach, a characteristic age for a sample of
galaxies replaces the detailed SFH of the individual galaxies in a SN survey.
Maoz et al. (2010) recently presented 
a method for recovering the DTD, which avoids this averaging.
In the method, the SFH of every individual galaxy, or even galaxy
subunit, is compared to the number of SNe it hosted in the survey
(generally none, sometimes one, rarely more). DTD recovery is treated
as a linearized inverse problem, which is solved statistically.
Maoz et al. (2010) applied the method to a subsample of the galaxies in the 
Lick Observatory SN Search (Filippenko et al. 2010; Leaman et
al. 2010; Li et al. 2010a,b), having SFH reconstructions  by Tojeiro et
al. (2009) based on data from the Sloan
Digital Sky Survey (SDSS; York et al, 2000). Maoz et al. (2010)
 find a significant detection of both
a prompt SN~Ia component, that explodes within 420~Myr of star formation, 
and a delayed SN~Ia with population that explodes after $>2.4$~Gyr . 
Maoz \& Badenes (2010) 
applied this method also to a sample of SN remnants in the Magellanic
Clouds, compiled by Badenes, Maoz, \& Draine (2010). Treating the remnants as
an effective SN survey conducted over $\sim 20$~kyr, they also find
a significant detection of a prompt (this time $<330$~Myr) SN~Ia
component.
A related DTD reconstruction method has been applied by Brandt et
al. (2010) to the SNe~Ia from the SDSS II survey. Like Maoz et
al. (2010), they detect both a prompt and a delayed SN~Ia population.

Yet another approach for recovering the DTD, which is
 at the focus of this
 paper, is to measure the SN rate
vs. redshift in massive galaxy clusters.
The deep potential wells of clusters, combined with their
relatively simple SFHs, make them ideal locations for
studying both the DTD and the metal production of SNe. 
Optical spectroscopy and multiwavelength photometry 
of cluster galaxies has shown consistently that the bulk of their
stars were formed within short episodes ($\sim 100$~Myr) 
at $z\sim 3$ (e.g., Daddi et al. 2000; Stanford et al. 2005;
 Eisenhardt et al. 2008). Thus, the
observed SN rate vs. the elapsed cosmic time 
since the stellar formation epoch 
provides an almost (see \S\ref{sscomparisontopred}) direct
measurement of the form of the DTD. 
Furthermore, the record of metals stored in   
the intracluster medium (ICM) constrains the  number of SNe that have
exploded, and hence the normalization of the DTD.

Renzini et
al. (1993) and Renzini  (1997) first pointed out that the large mass
of iron in the ICM could not have been produced solely by
CC SNe from
a stellar population with a standard initial mass function (IMF). 
Either a ``top-heavy'' IMF, or a
dominant contribution by SNe~Ia is required.  
Maoz \& Gal-Yam (2004) revisited the problem, and  tested the
hypothesis that the main source of iron is SNe~Ia. 
Accounting for the CC~SN contribution to the observed iron mass,
they calculated the number of SNe~Ia, per present-day stellar
luminosity, needed to have exploded over the entire past history of a
cluster, in order to produce the
observed iron mass. 
Finally, they compared some simple theoretical 
DTDs, normalized to produce this number of SNe, to
the then-available cluster SN~Ia rates (Gal-Yam, Maoz, \& Sharon 2002; 
Reiss 2000).
The low observed SN~Ia rates out to $z\sim 1$ implied that the large
number of events, needed to produce the bulk of the iron, occurred at
even higher redshifts, beyond the range of the then-existing observations. 
Maoz \& Gal-Yam (2004) therefore concluded that SNe~Ia can be the main source 
of iron only if most of them explode during the relatively brief time interval
between star formation in massive clusters (at $z\sim 2-3$) and the
highest-redshift cluster SN rate measurements (at $z\sim 1$). In other
words, the majority of SNe Ia (at least those that occur in a galaxy
cluster environment) must have a short time
delay, $\lesssim 2~$Gyr.

A shortcoming of the 
work by Maoz \& Gal-Yam (2004) was the paucity and quality of the
cluster SN rate data it was based on. The low-$z$ cluster SN rate by
Reiss (2000) was never published in the refereed literature. 
The cluster SN rates  at
$z=0.25-0.9$ by Gal-Yam et al. (2002), were based on very few
events -- two or three -- discovered in archival {\it Hubble
Space Telescope} (HST) WFPC2 images of clusters.
The large Poisson and systematic uncertainties
 in the resulting rates precluded
any detailed discrimination among model DTDs.
 
Over the past few years, the observational situation has improved dramatically.
Accurate  cluster SN~Ia rates have been measured in the redshift range
of 0 to 1.5, 
 with smaller errors than the
previous sole published study at $<\!\!z\!\!>=0.25, 0.90$ by Gal-Yam et
al. (2002). 
In this paper, we utilize this wealth of new measurements for a renewed
analysis of the cluster SN rate and its implications for the SN~Ia DTD
and for cluster metal-enrichment history. We show that the new data 
permit a direct comparison of the observations 
with the functional behavior of various DTDs that have been proposed in
the literature, whether parametrized-mathematical or phenomenological DTDs,
or DTDs resulting from  physical considerations at different levels of
sophistication.   
We find that few of these models, in their simplest forms,
 are compatible with the
emerging observational picture, and therefore we have reached the point where 
the observations discriminate among models.
 In \S2, below, we compile the existing 
cluster SN~Ia measurements.
In \S3, we review the existing measurements
of the properties of clusters, particularly the gas-to-stellar mass
ratio, that are relevant for estimating the present-day cluster
iron-to-stellar mass ratio. This ratio, in turn, sets the number of
SN~Ia that have exploded in clusters, per unit stellar mass formed,
which then fixes the normalization of the DTD. In \S4 we use the
rates and the normalization to recover the observational DTD.
We take a forward-modeling approach starting in \S5, where
we compare the observed rates to the DTD predictions of various 
``single-component'' models that invoke a single mathematical form
or physical progenitor formation channel. In \S\ref{ss.noninst}, we relax the
assumption of a single, instantaneous,
 star-formation episode in clusters, and examine
whether an extended star-formation history, combined with some single DTD,
can reproduce the measurements. In \S7, we examine the ability of 
composite models, which mix diverse DTDs, or multiple episodes
of star formation, to match the observations.
We summarize our results in \S8.
 Throughout the paper we assume a 
cosmology with parameters $\Omega_{\Lambda} = 0.7$, $\Omega_{m} =
0.3$, and $H_0 = 70$ km s$^{-1}$ Mpc$^{-1}$.

\section{Cluster SN~Ia rates}
\label{ssobsrates}
\subsection{Existing cluster SN rates}
To perform our analysis, we first compile the 
cluster SN~Ia rate measurements currently in existence.
The recent measurements are at mean redshifts of
$<\!\!z\!\!>=0.02$ (Mannucci et al 2008); $<\!\!z\!\!>=0.15$
(Sharon et al. 2007; Gal-Yam et al. 2008);
 $<\!\!z\!\!>=0.08, 0.23$ (Dilday et al. 2010); 
$<\!\!z\!\!>=0.46$ (Graham et al. 2008);
$<\!\!z\!\!>=0.60$ (Sharon et al. 2010); 
and $<\!\!z\!\!>=1.12$ (Barbary et al. 2010).
The mean redshifts are visibility-time-weighted
means for the clusters monitored by each survey, i.e., in the
calculation of the mean,
the redshift of
each cluster in the survey sample is weighted by the effective time
during which a SN~Ia would have been visible in the survey (see, e.g.,
Sharon et al. 2007). Before these recent measurements, the 
sole modern published cluster rates were
 at $<\!\!z\!\!>=0.25, 0.90$ (Gal-Yam et al. 2002). Table~1
 summarizes, for each of these measurements, the mean redshift of the
 sample, the redshift range, the cosmic times corresponding to these
redshifts in our assumed cosmology, 
 and the SN~Ia rate, normalized 
by stellar mass. The stellar masses, at the cosmic time
corresponding to the redshifts of the
 clusters, have generally 
been estimated 
consistently in the respective papers, based on the observed
 stellar luminosities and colors of the monitored cluster galaxies,
and assuming the same IMF -- the ``diet Salpeter'' IMF of Bell et al. (2003). 
Exceptions to this are: Dilday et al. (2010), who converted their
luminosity-normalized rates to mass-normalized rates assuming that the mean
$M/L_B$ ratio found by Sharon et al. (2007) 
 is valid for their cluster sample, which spans a similar redshift
 range; Gal-Yam et al. (2002), whose rates were mass normalized by 
Sharon et al. (2010), using $M/L_B$ from Sharon et al. (2007) at low
$z$ and $M/L_B$ from Sharon et al. (2010) at high
$z$; and Graham et al. (2008), who estimate masses for the galaxies in 
their sample using the spectral population synthesis
of Buzzoni (2005) who, in turn, assumed a Salpeter (1955) IMF. We therefore
scale up the Graham et al. (2008) rate by a factor of 1.77, 
which is
the ratio between the remaining stellar mass in an old and inactive quiescent 
population in a Salpeter (1955) IMF, and that in a diet-Salpeter IMF 
(see \S\ref{sscccont}, below, for details). We follow Barbary et
al. (2010) in scaling up the rate of Sharon et al. (2010) by 35\%, to
account for the offset in the $M/L_g$ vs. $g-r$ relation of
Bell et al. (2003), expected in the younger galaxies that exist 
at  $<\!\!z\!\!>=0.6$.

\begin{deluxetable*}{crcrrl}
\tablewidth{0pt} 
\tablecaption{{Cluster SN~Ia Rates and Delay Time Distribution}
\label{ratestable}}
\tabletypesize{\scriptsize}
\tablecolumns{6} 
\tablehead{\colhead{Redshift}    &
           \colhead{Cosmic time} &    
	    \colhead{SN~Ia rate} &    
	    \colhead{Delay}   	 &
	    \colhead{DTD}  &     
            \colhead{Ref.}\\	 
           \colhead{$z$}    &
           \colhead{$t$} &    
	    \colhead{$R_{Ia}(t)$} &    
	    \colhead{$\tau$}   	 &
	    \colhead{$\Psi(\tau)$}  &     
            \colhead{}\\	 
               \colhead{}    &
	  \colhead{[Gyr]}    &
	  \colhead{[SNuM]}    &
	  \colhead{[Gyr]}    &
	  \colhead{}	     &
           \colhead{}       \\   
\colhead{(1)}&
\colhead{(2)}&
\colhead{(3)}&
\colhead{(4)}&
\colhead{(5)}&
\colhead{(6)}\\
	  }
\startdata
 0.020$^{+0.020}_{-0.015}$&13.2$^{+0.2}_{-0.3}$&0.066$^{+0.027}_{-0.020}$&
11.1$^{+0.2 }_{-0.3}$ & 3.6$^{+1.5  }_{-1.1}$&M08\\
 0.084$^{+0.086}_{-0.054}$&12.4$^{+0.6}_{-1.0}$&0.060$^{+0.029}_{-0.021}$&
10.3$^{+0.6 }_{-1.0}$ & 3.3$^{+1.6  }_{-1.2}$&D10\\
 0.150$^{+0.040}_{-0.090}$&11.6$^{+1.0}_{-0.4}$&0.098$^{+0.068}_{-0.048}$&
 9.5$^{+1.0 }_{-0.4}$ & 5.5$^{+3.8  }_{-2.7}$&S07\\
 0.225$^{+0.075}_{-0.125}$&10.8$^{+1.3}_{-0.7}$&0.088$^{+0.025}_{-0.020}$&
 8.7$^{+1.3 }_{-0.7}$ & 4.9$^{+1.4  }_{-1.1}$&D10\\
 0.250$^{+0.120}_{-0.070}$&10.5$^{+0.7}_{-1.1}$&0.110$^{+0.160}_{-0.070}$&
 8.4$^{+0.7 }_{-1.1}$ & 6.2$^{+9.0  }_{-4.0}$&GY02\\
 0.460$^{+0.140}_{-0.260}$& 8.7$^{+ 2.3}_{-1.0}$&0.177$^{+0.212}_{-0.124}$&
 6.6$^{+2.3 }_{-1.0}$ &10.1$^{+12.1  }_{-7.1}$&G08\\
 0.600$^{+0.290}_{-0.100}$& 7.8$^{+ 0.7}_{-1.5}$&0.151$^{+0.138}_{-0.116}$&
 5.7$^{+0.7 }_{-1.5}$ & 8.7$^{+7.9  }_{-6.6}$&S10\\
 0.900$^{+0.370}_{-0.070}$& 6.2$^{+ 0.3}_{-1.4}$&0.220$^{+0.250}_{-0.110}$&
 4.1$^{+0.3 }_{-1.4}$ &12.9$^{+14.6  }_{-6.4}$&GY02\\
 1.120$^{+0.330}_{-0.220}$& 5.3$^{+ 0.9}_{-1.0}$&0.364$^{+0.301}_{-0.270}$&
 3.2$^{+0.9 }_{-1.0}$ &21.6$^{+19.9  }_{-16.0}$&B10\\
\nodata&\nodata&\nodata&1.1$^{+1.1 }_{-1.1}$ &230 $^{+112}_{-112}$&txt\\
\enddata

\tablenotetext{}{
(1) -- Visibility-time-weigthed 
mean redshift of cluster sample, and redshift range.\\
(2) -- Time since Big Bang corresponding to redshifts in (1).\\
(3) -- SN rate per unit stellar mass, in SNuM
($10^{-12}$~SNe~yr$^{-1} {\rm M_{\odot}}^{-1}$), with normalization relative
to remaining mass at the redshift of the observation.\\ 
(4) -- Time since $z_f=3$ corresponding to redshifts in (1).\\
(5) -- Mean recovered DTD value in time bin of (4), in units of 
$10^{-14}$~SNe~yr$^{-1} {\rm M_{\odot}}^{-1}$, with normalization relative 
to formed mass.\\ 
(6) -- References:
M08 -- Mannucci et al. (2008); D10 -- Dilday et al. (2010); 
S07 -- Sharon et al. (2007); GY02 -- Gal-Yam et al. (2002); 
G08 -- Graham et al. (2008); S10 -- Sharon et al. (2010); B10 -- Barbary et al. (2010); 
txt -- DTD value derived in this work, based on iron-mass constraints.\\
Published rates have been converted to the same ``diet Salpeter''
IMF (Bell et al. 2003). S10 rate has been scaled up to account for
expected evolution of the Bell et al. (2003) $M/L$-to-color relations,
according to B10.}
\end{deluxetable*}

\begin{figure}
\epsscale{1.23}
\plotone{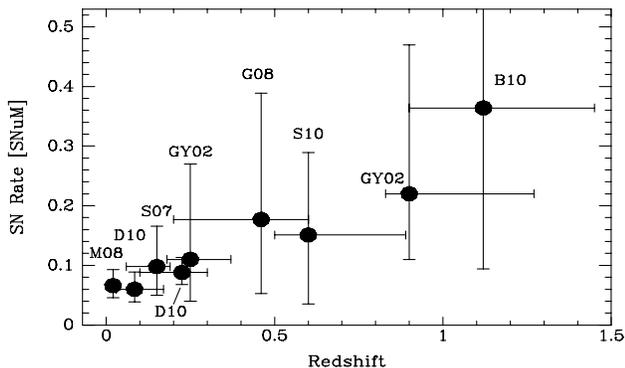}
\caption{SN~Ia rates per unit stellar mass in galaxy clusters, as 
a function of redshift, as listed in Table~1.
Observed rates, here and in the subsequent
Figs.~\ref{figplm0.5}$-$\ref{figmensfexpplusconst}, 
are in units of SNuM: $10^{-12}$~SNe~yr$^{-1} {\rm M_{\odot}}^{-1}$.
Horizontal error bars mark the cluster redshift intervals of the respective
SN surveys, with the central value at the visibility-time-weighted mean
redshift of each survey. Vertical error bars show the summed $68\%$
confidence limit
statistical and systematic uncertainties.
All measurements have been consistently scaled to the ``diet Salpeter'' IMF
(see text).
Labels are: M08 -- Mannucci et al. (2008); D10 -- Dilday et al. (2010); 
S07 -- Sharon et al. (2007); GY02 -- Gal-Yam et al. (2002); 
G08 -- Graham et al. (2008); S10 -- Sharon et al. (2010); B10 -- Barbary et
al. (2010).
} 
\label{figobsrates}.
\end{figure}

\subsection{Observed cluster SN~Ia rates vs. redshift} 
Figure~\ref{figobsrates} shows the rates we have compiled, and which we analyze
in this paper. It suggests only a mild rise in the rates,
by a factor of $\sim 2-5$, over the redshift interval probed. 
Remarkably, the cluster SN rate
appears not to change much from $z=0$ to $z=1.45$, corresponding to a
lookback time of 9.8~Gyr. This is only about 2~Gyr after
the formation of the cluster stars, assuming this occurred
 at a redshift $z_f\approx 3$. 

\section{Time-integrated SN number per formed mass}
\label{ssnoverm}
As pointed out by Maoz \& Gal-Yam (2004) and 
Sharon et al. (2010), and now further suggested by
the results above, the integral over the SN~Ia rate from $z=0$ to $z=1.45$,
times the mean iron yield of SNe~Ia ($0.6-0.7 {\rm M_\odot}$; see below), 
gives just 
a small fraction, roughly 10\%, of the present-day ratio of iron mass
to stellar mass observed in clusters. 
It is then unavoidable
that the large majority of cluster metals were produced within the
first 1-2~Gyrs after the formation of the cluster stars. The metals were
produced by CC SNe and perhaps by
``prompt'' SNe~Ia, 
but they were not produced by any SNe with a larger delay, as the rate 
measurements show that only few such explosions took place
at those later times. 

In order to 
juxtapose the new cluster rates, compiled above, with model expectations,
we revisit the question of the 
iron-to-stellar mass ratio in clusters, which dictates the
normalization
of the DTD. In addition to  
considering the latest data on cluster
properties, our treatment will differ from the one in Maoz \& Gal-Yam
(2004) in several respects, most notably in that we will use
mass-normalized, rather than luminosity-normalized SN rates, and we
will account for mass evolution due to stellar mass loss.
Stellar mass is, of course, more closely related to the number of stars 
than is blue luminosity, and is a more stable quantity.

The mass of iron in clusters that is attributable to SNe~Ia is the
sum of the iron masses in the ICM and in stars, minus the mass of iron
produced by CC~SNe. Thus, the ratio between SN~Ia-produced iron mass and
the present-day stellar mass is
\begin{equation}
\frac{M_{\rm Fe, Ia}}{M_{*,0}}=0.0074[
\frac{Z_{{\rm Fe},\odot}}{0.0026}\left(\frac{M_{\rm gas}/M_{*,0}}{10}\frac 
   {Z_{\rm Fe, gas}}{0.3}+\frac{Z_{\rm Fe,*}}{1.2}\right) 
\label{irontostar}
\end{equation}
$$
   -\frac{M(>8{\rm M_\odot})/M_{*,0}}{0.35}\frac{y_{\rm cc}}{0.01}].
$$
We discuss below each of the quantities that enter this equation.

\subsection{The solar iron abundance}
The value of the photospheric solar abundance of iron,
$Z_{{\rm Fe},\odot}=0.0026$, is 
from Anders and Grevesse (1989). Although the correct photospheric iron
abundance has been revised down to 0.00176 (Grevesse \& Sauval 1999), 
the ICM and stellar abundances we quote
below, and which are still generally used in the literature, relate to
the Anders and Grevesse (1989) value\footnote{Lin et al. (2003) derived
an iron-to-stellar mass ratio, using a fairly low ICM iron abundance of
0.21 solar found by De Grandi \& Molendi (2001), 
but 
coupled it to the {\it meteoritic} iron abundance
of Anders and Grevesse (1989), 0.00181. As a result, the ratio they found
is a factor of $\sim 2$ lower than we find here.}.

\subsection{The gas-to-star mass ratio}
\label{ssgastostar}
The present-day 
mass ratio of baryons in the ICM gas and in stars, $M_{\rm gas}/M_{*,0}$,
has been  recently re-evaluated by Gonzalez, Zaritsky, \& Zabludoff
(2007), taking into account the contribution to $M_{*,0}$ by the intergalactic
stellar population of a cluster. 
Contrary to Lin et
al. (2003), who argued for $M_{\rm gas}/M_{*,0}$ that is independent of
total cluster mass,
Gonzalez et al. (2007) find that $M_{\rm
  gas}/M_{*,0}$ rises monotonically with cluster mass, from a value
of about 5 for low-mass clusters to about 15 for the most-massive
systems. To find the suitable values for 
the clusters from which were
derived the SN rates that we analyze, we now examine the typical
masses of those clusters.
 
The most massive cluster sample in our SN rate compilation is
that of Sharon et al. (2010), which is largely coincident with the 
 $z>0.5$ MACS cluster sample of Ebeling et
al. (2007). The clusters in this sample have total masses 
$M_{500}\sim 10^{15} {\rm M_\odot}$, where $M_{500}$ is
the mass within a spherical volume in which the mean density is 500 times
the critical closure density.
In the
Sharon et al. (2010) sample, the part of the rest-frame $B$-band
luminosity of each cluster that is included in the HST field of view
(and which is therefore monitored for SNe)
is $(1-6)\times 10^{12} L_{B,\odot}$, and the corresponding
stellar 
mass range is $(4-17)\times 10^{12} {\rm M}_\odot$, with a mean 
of $10^{13} {\rm M}_\odot$.  
Such masses also apply to the massive, X-ray-selected clusters 
at $<\!\!z\!\!>=0.25, 0.90$
studied
by Gal-Yam et al. (2002).
For the $<\!\!z\!\!>=0.15$ 
clusters monitored by Gal-Yam et al. (2008),
the typical stellar 
mass is $7\times 10^{12} {\rm M_\odot}$ (Sharon et al. 2007).
From Gonzalez et
al. (2007), this
corresponds to total masses
of  $M_{500}\sim 4\times 10^{14} {\rm M_\odot}$.

Barbary et al. (2010) find, for their $<\!\!z\!\!>=1.12$ clusters,
a mean $B$-band
luminosity, of 
 $\approx 3 \times 10^{12} L_{B,\odot}$, and typical stellar masses
of $\approx 4\times 10^{12} {\rm M_\odot}$. As a check,
the bulk properties of the Barbary et al. (2010) clusters, 
such as X-ray luminosities, X-ray temperatures, and velocity 
dispersions, also suggest masses about one-half as large 
as those of the cluster sample of Sharon et al. (2010) at
$<\!\!z\!\!>=0.60$. 
Well-measured X-ray temperatures exist in the literature
for five of the Barbary et al. sample clusters, and are all around
6-7 keV (Rosati et al. 2004; 
Stanford et al. 2002, 2006; Boehringer et al. 2008; Gilbank et
al. 2008). 
This compares to the mean temperature,
$<\!\!kT\!\!>=9$~keV, of the $z>0.5$ MACS cluster sample of Ebeling et
al. (2007). X-ray luminosities, reported for these five Barbary et al. 
clusters, and for 
an additional four of their clusters
(Rosati et al. 1999; Postman
et al. 2001; Bremer et al. 2006; Andreon et al. 2008) are in
the range $\sim 0.7-16\times 10^{44}~{\rm erg~s}^{-1}$, 
with a mean of $\sim 4\times 10^{44}~{\rm erg~s}^{-1}$,
compared to
$<\!\!L_X\!\!>=16\times 10^{44}~{\rm erg~s}^{-1}$ for the Ebeling et
al. (2007) sample. Galaxy velocity dispersion best-fit measurements
for seven of the Barbary et al. clusters are in the range of $\approx
600-1300~{\rm km~s}^{-1}$ (references above, plus Eisenhardt et
al. 2008; Cain et al. 2008), with a mean of about $850~{\rm
  km~s}^{-1}$, compared to $<\!\!\sigma\!\!>=1300~{\rm km~s}^{-1}$
for the $z>0.5$ MACS clusters. 
Since total mass depends (e.g., Hicks et al. 2008) 
on X-ray temperature roughly as $M\propto
T^{1.5}$, on X-ray luminosity roughly as $M\propto
L_X^{0.5}$, and on velocity dispersion as $M\propto
\sigma^2$, all of these observables suggest that the typical cluster
in the Barbary et al. sample has about one-half the mass of the typical
cluster in the Sharon et al. (2010) sample.
Assuming a constant fraction of the cluster mass in stars, the stellar
mass in the Barbary et al. sample would also be of order one-half that 
in the Sharon et al. (2010) sample. However, the stellar mass fraction
in clusters depends on total cluster mass. For example, Andreon (2010)
has recently found a dependence of roughly $f_{\rm stars}\propto
M^{-0.5}$. If this relation held also for clusters at redshifts $z>1$, 
the stellar mass in the Barbary et al. clusters would be only about 1.4 times
lower than in the Sharon et al. (2010) sample, and thus more similar
to the $<\!\!z\!\!>=0.15$ 
clusters monitored by Gal-Yam et al. (2008),

The optically selected clusters monitored for SNe by Graham 
et al. (2008) and Dilday et al. (2010) have masses comparable
to those of the samples discussed above, though somewhat lower. 
Graham et al. (2008) derived a total
stellar mass of $M_*=8\times 10^{13}{\rm M_\odot}$ for the 30 clusters
they used for their SN rate measurement, or $\sim 3\times 10^{12}{\rm M_\odot}$
for a typical cluster. Dilday et al. (2010) report a total $r$-band
luminosity of $L_r=10^{14}{\rm L_{r,\odot}}$ for the 71 clusters in their
low-$z$, ``C4'', cluster subsample. Taking $M_*/L_r=3$ (Sharon et
al. 2007), this gives a mean stellar mass per cluster of 
$\sim 4\times 10^{12}{\rm M_\odot}$. For the 492 clusters in the
$<\!\!z\!\!>=0.225$ ``maxBCG'' sample of Dilday et al. (2010), the integrated
luminosity is $L_r=2\times 10^{14}{\rm L_{r,\odot}}$, and hence
the mean stellar mass per cluster is $\sim 1\times 10^{12}{\rm M_\odot}$. 
Thus, this one subsample of Dilday et al. (2010) has clusters that are
undermassive, compared to the other cluster samples we
analyze. The other samples have only about a factor-3 difference in mean
cluster mass between the higher- and lower-mass extremes.

 From the fits by Gonzalez et al. (2007), the lower and higher
typical  masses considered above (excluding the lower-mass maxBCG sample of
Dilday et al. 2010) correspond to  
$M_{\rm gas}/M_{*,0}$ values of 6.5 and 14.5, respectively. However,
Gonzalez et al. (2007) used the relation obtained by Cappellari et
al. (2006) between the kinematically measured
total-mass-to-luminosity ratio, $M/L$  of nearby ellipticals and
their $I$-band luminosity. Cappellari et al., when 
comparing their results to $M/L$
estimates for the same galaxies based on spectral population 
synthesis,  concluded that 30\% of the contribution 
to their kinematic $M/L$ could be due to dark matter within the central
regions. If so, the stellar masses in clusters found by Gonzalez et
al. (2007) would be scaled down by 0.7, and hence
$M_{\rm gas}/M_{*,0}$ would increase to 9.3 for, e.g., the
$<\!\!z\!\!>=0.15$ cluster sample, 
 and to 21 for the $<\!\!z\!\!>=0.60$ cluster sample.

Lagan\'a et al. (2008) re-analyzed X-ray and optical data for five
clusters to deduce the stellar and gas mass fractions. To obtain
stellar masses, they used the Kauffmann et al. (2003) $M/L$ ratios,
which are consistent with the Bell et al. (2003) $M/L$ ratios, used
in deriving the SN rates analyzed here. For clusters with total masses, 
$M_{500}$, 
similar to the two ends of the typical 
mass range considered here, Lagan\'a et al. (2008)  
find $M_{\rm gas}/M_{*,0}$ of about 8 and 17, respectively.  
Giodini et
al. (2009) also recently examined the dependence of stellar and gas
mass fractions on $M_{500}$. Gas mass fractions were estimated
for 41 clusters compiled by Pratt et al. (2009) from the literature.
Stellar mass fractions were obtained
for a sample of 91 groups and poor clusters from the COSMOS survey, 
based on spectral population synthesis fitting of their spectral
energy distributions. These were combined with the re-evaluated 
stellar masses of 27 clusters from Lin et al. (2003). We scale down
their stellar masses by 0.7, to transform from the Salpeter (1955) IMF
they assumed to our assumed 
``diet Salpeter'' IMF (see below). For the two ends of the cluster mass range
considered above, the best-fit relations of Giodini et
al. (2009) then give  $M_{\rm gas}/M_{*,0}$ of about 7 and 11, respectively.  
Finally, a recent analysis by Andreon (2010) of 52 clusters, chosen to have
accurate measurements, results in similar relations, with values
of $M_{\rm gas}/M_{*,0}$ of about 7 and 13.5 for the two typical
cluster 
masses, respectively. 
  
Considering all of the above studies, we see that the estimates of 
$M_{\rm gas}/M_{*,0}$ are in the 
range of $6-10$ for the lower-mass clusters of the type monitored 
by most of the low-$z$ surveys whose SN rates we analyze, 
$11-21$ for the
massive, intermediate-$z$ 
clusters monitored by Gal-Yam et al. (2002) and by Sharon et
al. (2010), and somewhere inside this range for the high-$z$
cluster sample of Barbary et al. (2010).
 (We also note that these
ratios appear to be independent of redshift; e.g. Giodini et al. 2009).
When
considering the iron mass in clusters and the number of SNe~Ia needed
to produce it, we therefore
choose $M_{\rm gas}/M_{*,0}=10$ as our ``optimal'' fiducial value, 
but we will also consider the consequences of a ``minimal iron'' value
of $M_{\rm gas}/M_{*,0}=6$.   

\subsection{The cluster iron adundance}
A ``canonical'' value of $Z_{\rm
  Fe, gas}=0.3$ is by now fairly well-established
 for the ICM gas iron abundance, relative
to solar. Balestra et al. (2006), Maughan et al. (2008), and Anderson
  et al. (2009) argue for
values that have evolved even higher, to $0.4-0.5$, at low redshifts
(but see Ehlert \& Ulmer 2009). 
 To err conservatively, we 
adopt $Z_{\rm  Fe, gas}=0.3$ for out fiducial value. These iron abundances
typically relate to the central $\sim 1$~Mpc radius of massive clusters, the
  same region in which the stellar luminosity is measured and the SNe
  are detected by the cluster SN surveys we consider. Nevertheless,
 because
the X-ray emission is strongly centrally peaked, there is some concern
that the above iron abundance could be strongly biased by a high 
abundance in the very core region, which might be unrepresentatative of
  a much lower abundance existing over  most of the volume within 1~Mpc.
However, Maughan et al. (2008) have shown that measuring iron
  abundances within an annulus that excludes the central 15\% in radius  
reduces the results only mildly, to values still consistent with the
canonical $Z_{\rm  Fe, gas}=0.3$.  
For the stellar iron abundance relative to solar, we follow Maoz \&
Gal-Yam (2004) and Lin et al. (2003) in adopting the value 
$Z_{\rm Fe,*}=1.2$ from the study by Jorgensen et al. (1996).

\subsection{The CC~SN contribution}
\label{sscccont}
$M(>8{\rm M_\odot})/M_{*,0}$ is the ratio between the initial mass in stars of
mass above $8{\rm M_\odot}$, which will explode as CC SNe, and the
present-day mass in stars in clusters. Maoz \& Gal-Yam (2004)
examined the dependence of this ratio on the choice of 
IMF for a large range of standard and
non-standard IMFs. In the present work, we will consistently assume the
``diet Salpeter'' IMF of Bell et al. (2003), as this is the IMF that
has been assumed when determining rates per unit mass in the SN
surveys that we analyze. This IMF is like the Salpeter (1955)
single-power-law IMF with slope $-2.35$ between 0.1 and 100${\rm
  M_\odot}$. However, when calculating observables involving mass,
such as $M/L$,
the total initial mass in stars is scaled
down by a factor of $0.7$, to simulate the deficit, relative to the
Salpeter IMF, of low-mass stars in realistic IMFs. 
The ratio of  mass in stars of
mass above $8{\rm M_\odot}$ to the total initial mass, 
in such an IMF, is 0.20.
 From Bruzual \& Charlot (2003), for a Salpeter
(1955) IMF, during the stellar evolution of
a 10~Gyr-old population 31\% of the stellar mass is returned to the
interstellar medium (ISM) via stellar winds and SN explosions. For a diet
Salpeter IMF, this mass-loss fraction is $0.31/0.7=44\%$, and hence
the present-day ratio $M(>8{\rm M_\odot})/M_{*,0}\approx 0.2/0.56=0.35$.
This is, in the present context, a conservative estimate, in that it 
maximizes the contribution of CC~SNe, relative to SNe~Ia, 
to cluster iron production. If, as generally believed, stars in some mass
ranges collapse directly into black holes, without a SN explosion and its
contribution to iron production,
then the appropriate mass ratio to be used here would be lower. For
example, if only stars up to $50{\rm M_\odot}$ or $25 {\rm M_\odot}$ explode as CC
SNe, $M(>8{\rm M_\odot})/M_{*,0}$ will be reduced to $0.29$ or $0.20$, 
respectively. 
 
The diet Salpeter IMF gives a very
similar $M(>8{\rm M_\odot})/M_{*,0}$ ratio 
to that of the Kroupa (2001) and Gould et al. (1997)
IMFs (see Maoz \& Gal-Yam 2004 and Bell et al. 2003). In any case,
since both the SN rate and the iron mass in clusters are 
 normalized relative to stellar mass, as derived from observed stellar
 luminosity, the conclusions are insensitive to the assumed
 form of the IMF in the low-mass range, as long as it is assumed consistently
 for both. The ratio $M(>8{\rm M_\odot})/M_{*,0}$ also depends weakly on the
 assumed age of the population, as long as it is of order several Gyr
 or more, since most of the return of mass to the ISM occurs early on.
 
Finally, we follow Maoz \& Gal-Yam (2004) in assuming that the iron
yield of a CC SN, $y_{\rm cc}$, is 1\% of the initial mass of the
progenitor star. As discussed there, this is likely to be a
conservative overestimate as well. For $M(>8{\rm M_\odot})/M_{*,0}=0.35$ and the
fiducial values of the metallicities, this means about 1/3 of the
cluster iron mass was produced by CC~SNe. For
$M(>8{\rm M_\odot})/M_{*,0}=0.20$, less than 1/5 of the
iron mass is from CC~SNe. Alternatively, for the ``minimal iron'' value
of $M_{\rm gas}/M_{*,0}=6$, the core-collapse contribution is
between 1/2 and 1/4.

\subsection{The SN~Ia iron yield and correction for stellar mass evolution}
Dividing Eq.~\ref{irontostar} by the mean iron yield of SNe~Ia gives 
the time-integrated number of SNe~Ia that have
exploded in clusters, per present-day unit stellar mass, $N_{\rm
  SN}/M_{*,0}$. 
The mean iron yield of a SN~Ia is  $0.6-0.7~{\rm M_\odot}$ (e.g., Mazzali
et al. 2007). In line with our conservative approach, we will assume
 the higher value of $0.7~{\rm M_\odot}$, thus lowering the 
integrated number of SNe~Ia  required to explain the observed abundances.

The time-integrated number of SNe~Ia per unit {\it formed}
stellar mass, before the 44\% mass loss that occurs over the course of $\sim
10$~Gyr (see \S\ref{sscccont}, above), will be 0.56 times lower.
Multiplying Eq.~\ref{irontostar} by 
$0.56/0.7 {\rm M_\odot}$, we thus obtain the time-integrated 
number of cluster SNe~Ia per unit formed stellar mass,
\begin{equation}
\left(\frac{N_{\rm SN}}{M_*}\right)_{\rm opt}
=0.0059~ {\rm M_\odot}^{-1}=5.9~{\rm SNuM ~Gyr}, 
\label{eqoptiron}
\end{equation}
assuming the ``optimal'' value of $M_{\rm gas}/M_{*,0}=10$, and the
other fiducial values in Eq.~\ref{irontostar}. Alternatively, 
for the ``minimal'' value of $M_{\rm gas}/M_{*,0}=6$, we obtain
\begin{equation}
\left(\frac{N_{\rm SN}}{M_*}\right)_{\rm min}=0.0034~ {\rm M_\odot}^{-1}=3.4~{\rm SNuM ~Gyr}, 
\label{eqminiron}
\end{equation}

\section{A reconstruction of the observational SN delay-time distribution}
\label{ssdtd}
In this section, we combine the observed cluster SN rates, $R_{Ia}(t)$
(from \S\ref{ssobsrates})
and the integrated SN~Ia number per formed stellar mass, $N_{\rm
  SN}/M_*$ (from \S\ref{ssnoverm}),
to recover the SN~Ia DTD. 
In the next section, we take an alternative, forward-modeling,
approach, to examine the predictions of a variety
of specific model DTDs.  
Throughout this section, 
we assume that all stars in clusters formed in a single instantaneous
burst, neglecting any subsequent star formation. As already noted above, this 
is a fair approximation to the results of spectral synthesis of the 
mass-dominant stellar populations in cluster ellipticals, indicating
 a burst lasting $\sim 0.1$~Gyr
at $z_f\approx 3$ (corresponding to cosmic time $t_f\approx
2$~Gyr), i.e., some 11.5~Gyr ago.
The consequences of relaxing the instantaneous-burst
assumption are studied in \S\ref{ss.noninst}-\ref{ssdoublecomp}, below.
 If, in fact, the burst assumption is valid,
recovery of the DTD is straightforward.
The SN rate since the epoch of star formation
differs from the DTD only in the stellar-mass normalization -- the DTD
is the SN rate normalized by unit stellar mass at the formation epoch, $M_*$,
while the SN rate is
normalized by the remaining stellar mass, $M_*(t)=M_*~m(t)$, 
at the cosmic epoch of the 
rate measurement. Here, $m(t)$ is the remaining fraction of the initially
formed stellar mass at cosmic time $t$ (since the Big Bang). 
The observed SN rates
per unit mass at cosmic times $t$ are the values of the DTD at delays
$t-t_f$, up to the correction, easily applied, to convert from the
existing mass at time $t$ to the formed mass at $t_f$, accounting for
mass loss during stellar evolution.
The SN rate 
and the DTD are thus related by 
\begin{equation}
R_{Ia}(t)=\frac{\Psi(t-t_f)}{m(t-t_f)}.
\label{ddtmassloss}
\end{equation}

Bruzual \& Charlot (2003) tabulate the relative accumulated stellar
mass loss versus time, 
$m_{\rm loss}(t)$, following a burst of star formation, assuming a Salpeter
(1955) IMF. For the ``diet Salpeter'' IMF, the remaining mass fraction is
   $m(t)=1-m_{\rm loss}(t)/0.7$. Starting 2.5~Myr after star
formation, the remaining mass $m(t)$ is well approximated as $(t/2.5~{\rm
  Myr})^{-0.07}$. The predicted SN rate will therefore decline slightly
less steeply than the DTD. We convert the observed SN rates,
$R_{Ia}$, to the DTD, $\Psi(t)$,  using Eq.~\ref{ddtmassloss} with
the full Bruzual \& Charlot (2003) mass loss (rather than the
power-law approximation).    

 To estimate the DTD at delay times
earlier than those corresponding to the redshift, $z_{\rm max}$, of
the most distant clusters
monitored, we can use the $N_{\rm SN}/M_*$ constraints from the iron
observations. The total number of SNe per unit formed stellar mass
between $z_f$ and $z_{\rm max}$ is $(N_{\rm SN}/M_*)_{z_f,z_{\rm max}}
=(N_{\rm SN}/M_*)-(N_{\rm SN}/M_*)_{z_{\rm max, 0}}$, where
$(N_{\rm SN}/M_*)_{z_{\rm max, 0}}$ is the total number between $z=0$
    and $z_{\rm max}$, obtained by integrating the DTD over that
    range. The mean DTD value in the time interval $[0, t_{\rm max}-t_f]$, 
corresponding to the
 unobserved redshift range $z_f$ to $z_{\rm max}$, is just  
$(N_{\rm SN}/M_*)_{z_f,z_{\rm max}}/(t_{\rm max}-t_f)$.  

 To estimate $(N_{\rm SN}/M_*)_{z_{\rm max, 0}}$, we integrate, using
 small time steps, over the mass-loss-corrected SN rates. At times that
 are covered by several measurements, we take a mean rate, weighted by
 the relative measurement errors. Out to $z=1.45$, we find an
 integrated SN to stellar mass ratio of $(N_{\rm SN}/M_*)_{z_{\rm max,
 0}}=8.2\times 10^{-4}~{\rm M_\odot}^{-1}$. 
Subtracting this from the minimal-iron value of
 $0.0034~{\rm M_\odot}^{-1}$ (Eq.~\ref{eqminiron}) leaves $0.0026~{\rm M_\odot}^{-1}$ in the time 
interval of 
 2.2~Gyr between  $z_f=3$ to $z_{\rm max}=1.45$, or a mean DTD value of 
0.012~SNe~yr$^{-1}(10^{10}{\rm M_\odot})^{-1}$ in this delay time interval
 of $0-2.2$~Gyr.
 For the optimal-iron value (Eq.~\ref{eqoptiron}) of
 $0.0059~{\rm M_\odot}^{-1}$, the mean DTD value in this time bin is 
0.023~SNe~yr$^{-1}(10^{10}{\rm M_\odot})^{-1}$. We take this range of
DTD values as the uncertainty, which is driven in this case
mainly by the systematic uncertainty in the cluster gas-to-stellar
 mass ratio. For our best estimate we take the optimal-iron value.
We note that this exercise implies that, for the minimal and optimal
 iron assumptions, respectively, 79\% to 88\% of the SNe~Ia
 in clusters exploded before $z=1.45$.
\begin{figure}
\epsscale{1.23}
\plotone{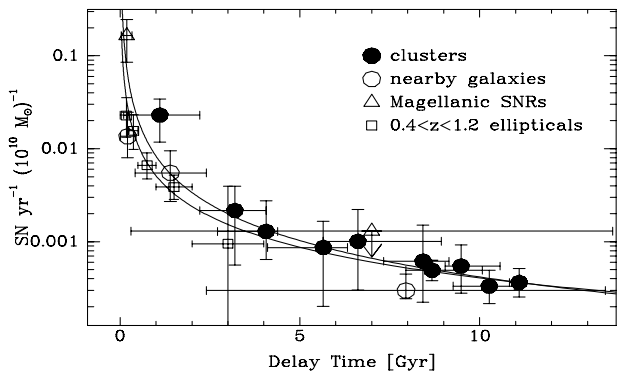}
\caption{Recovered SN~Ia 
delay-time distribution (DTD) for galaxy clusters (filled
circles). All but the earliest point are obtained directly from the  
observed rates in Fig.~\ref{figobsrates}, after correcting them for
stellar mass loss, and assuming an instantaneous star-formation burst at 
$z_f=3$ for all clusters. Horizontal error bars indicate the time bin
over which each mean DTD value is evaluated.
The first point is obtained by requiring a
time-integrated number of SNe~Ia according to 
Eqns.~\ref{eqoptiron}-\ref{eqminiron}, satisfying the observed iron-to-stellar 
mass ratio in clusters. Also plotted, for comparison, are the recovered
observational
DTDs from: Maoz et al. (2010), based on SNe in nearby galaxies from LOSS 
(empty circles); 
Maoz \& Badenes (2010), based on SN remnants in the Magellanic Clouds
(empty triangles, with arrow marking the 95\% confidence upper limit);
and Totani et al. (2008) based on SN~Ia candidates in elliptical galaxies
at $z=0.4-1.2$ (empty squares). 
For reference, we also show (curves) 
power laws, $t^{-s}$, with $s=-1.1$ and $s=-1.3$, scaled
to pass through the latest cluster-based point.} 
\label{figdtd}.
\end{figure}

Figure~\ref{figdtd} shows the recovered
 DTD, whose values are also listed in Table 1. 
The horizontal error bars mark the limits of each DTD time bin,
while the vertical error bars show the 
summed (not in quadrature) and propagated 
statistical and sytematic errors.
Also shown in 
Fig.~\ref{figdtd} are three other recently recovered observational
SN~Ia DTDs. One (empty circles) is from Maoz et al. (2010), obtained 
for a subsample
 of the local
 galaxies in the Lick Observatory SN Search having individual 
SFH reconstructions
based on SDSS spectra. We have plotted here the Maoz et al. (2010) 
DTD from a subsample
that excluded galaxies of Hubble types Sa to Sbc, in order to reduce
cross talk between time bins. Due to the limited aperture of the SDSS 
spectrograph fibers, there remains
 a ``leak'' of signal from the first to the second time bins, and therefore
the plotted DTD level in the $40-420$~Myr time bin is likely an underestimate 
of the true level (see Maoz et al. 2010). 
Another DTD shown (empty triangles) was found 
by Maoz \& Badenes
 (2010), using the SN remnants in the Magellanic Clouds as an
 effective SN survey. In this DTD, there is a detection 
in one time bin, of $t<330$~Myr, but only an upper limit on the
DTD value between 330~Myr to a Hubble time, which is indicated. Finally,
we show (empty squares) the DTD  recovered by 
Totani et al. (2008), using SN~Ia candidates in 
SXDS field elliptical galaxies at $z=0.4-1.2$. 
We have divided the DTD values of 
Totani et al. by a factor $1.8$, given by them,
 to convert from their normalization unit, of present-day $K$-band luminosity, 
to ours, of formed stellar mass. We have then divided by a further factor 0.7
to convert from their assumed Salpeter (1955) IMF to our diet-Salpeter IMF.
We also show, for reference, two power laws, $\Psi\propto t^{-1.1}$ and
$\Psi\propto t^{-1.3}$, which are discussed further in
 \S~\ref{sspowerlawdtd}, below. 

Several facts are apparent in Fig.~\ref{figdtd}.
First, considering the different time bins of the different measurements,
and the known systematics, the various measurements are generally consistent
with each other in regions of overlap. Second, the DTD is a monotonically
decreasing function, peaking at the earliest delays probed by the observations.
Finally all the plotted points appear to be generally 
 consistent with the illustrated power-law
dependences, which, remarkably,
 pass through each and every error box. This consistency 
is tested quantitatively in \S~\ref{sspowerlawdtd}.
With this new DTD derivation, combined with
those by Maoz et al. (2010) and Maoz \& Badenes (2010), we thus confirm
and extend the results of Totani et al. (2008), who found 
in their field elliptical SN sample a best-fit DTD dependence of a power law,
$\Psi\propto t^{s}$, with $s=1.08\pm 0.15$.
 This is also further discussed below,
in \S\ref{sspowerlawdtd}.

\section{Comparison to DTD models -- Single-component models}
\label{sscomparisontopred}

In this section, we take a forward-modeling
approach, to compare the predictions of specific model DTDs to the 
observations.
Foward modeling is advantageous in that the actual data are not
manipulated and errors need not be propagated. Furthermore, 
this approach will permit us later
(\S\ref{ss.noninst}-\ref{ssdoublecomp}) 
to relax the assumption of a single short
burst of star formation in clusters.

A model DTD that we test, $\Psi(t)$,  in order to be consistent with 
observed cluster iron abundances,  needs to have  a normalization such that
its integral over a cluster stellar age, $t_0$, 
agrees with the above, time-integrated, SN numbers:
\begin{equation}
\int_0^{t_0} \Psi (t) dt=\frac{N_{\rm SN}}{M_*}.
\label{normconstraint}
\end{equation} 
This is one constraint on the DTD, imposed by the cluster iron
abundances. A second set of constraints is imposed by the observed 
 time dependence of the SN rate, which can be compared to the 
model DTD predictions. The predictions will be tested with the
$\chi^2$ figure of merit. The SN rate at each visibility-time-weighted 
mean redshift is compared with the model prediction, with the latter
averaged over the
redshift interval spanned by the cluster sample used for deriving each
SN rate. Errors on the observed rates are generally asymmetric. For the
$\chi^2$ calculation, we use the error that is on the model side of each 
data point.

Throughout this section, 
we  assume that all stars in clusters formed in a single instantaneous
burst, neglecting any subsequent star formation. As already noted above, this 
is a fair approximation to the results of spectral synthesis of the 
mass-dominant stellar populations in cluster ellipticals, indicating
 a short
burst at $z_f\approx 3$.
As noted in \S\ref{ssdtd},
the SN rate since the epoch of star formation
differs from the DTD only in the stellar-mass normalization, whether 
formed mass, $M_*$,
or remaining stellar mass, $M_*(t)=M_*~m(t)$, 
at the epoch of the 
rate measurement, and we again use Eq.~\ref{ddtmassloss},
this time to convert the DTD model, $\Psi(t)$,
 to a rate prediction, $R_{Ia}(t)$.
 
Various forms have been proposed for the DTD, some derived from detailed
binary population synthesis calculations
(e.g., Yungelson \& Livio 2000;  Han \& Podsiadlowski
2004; Ruiter et al. 2009; Bogomazov \& Tutukov et al. 2009; 
Mennekens et al. 2010);
some physically motivated mathematical parameterizations, with varying
degrees of sophistication (e.g., Madau
et al. 1998; Greggio 2005; Totani et al. 2008); 
and some ad hoc formulations intended to
reproduce the observed field SN rate evolution (e.g., Strolger et
al. 2004).  We now attempt to fit the observed cluster SN~Ia
rates with some of these proposed DTDs, using Eq.~\ref{ddtmassloss},
while simultaneously
satisfying the DTD normalization constraint, Eq.~\ref{normconstraint},
set by the iron abundances in clusters. 

\subsection{Power-law DTDs}
\label{sspowerlawdtd}
A first and  simple mathematical parametrization of the DTD that we 
 compare to the observed cluster SN rates is a  
power law in time, $\Psi(t)\propto t^s$. Power laws
 have been long considered as possible
forms of the DTD (e.g., Ciotti et al. 1991; Sadat et
 al. 1998). As noted by previous authors (e.g., Greggio 2005; 
Totani et al. 2008) a power-law dependence 
is generic to models (such as the DD model) in which 
the event rate ultimately depends on the loss of energy and angular momentum 
to gravitational radiation by the progenitor binary system. 
If the dynamics are controlled solely by gravitational wave losses,
the time $t$ until a merger depends on the
binary separation $a$ as
\begin{equation}
t\sim a^4.
\end{equation}
If the separations are distributed as a power law
\begin{equation}
\frac{dN}{da}\sim a^\epsilon,
\end{equation}
then the event rate will be
\begin{equation}
\frac{dN}{dt}=\frac{dN}{da}\frac{da}{dt}\sim t^{(\epsilon -3)/4} .
\label{DDdependence}
\end{equation}
For  a fairly large range around $\epsilon\approx -1$, which describes
well the observed distribution of initial separations  of non-interacting
binaries (see Maoz 2008 for a review of the issue in the present context),
the DTD will
have a power-law dependence with index not far from $-1$.
However, in
reality, the binary separation distribution of WDs that have
emerged from their common envelope phase could be radically different,
given the complexity of the physics of that phase. Thus, the $\sim
t^{-1}$ DTD dependence of the DD channel cannot be considered unavoidable. 
Be that as it may, the DTD reconstructions by Totani et al. (2008),
Maoz et al. (2010), Maoz \& Badenes (2010), and in this work
(\S\ref{ssdtd}), all point to a $\sim t^{-1}$ power-law. 

A different power-law DTD dependence, with different physical
 motivation, has
been proposed by Pritchet et al. (2008). If the time between formation
of a WD and its explosion as a SN~Ia is always brief compared to the
formation time of the WD, the DTD will simply be proportional to the 
formation rate of WDs. Assuming that the main-sequence lifetime of a star
depends on its initial mass, $m$, as  a power law,
\begin{equation}
t\sim m^\delta,
\end{equation}
and assuming the IMF is also a power law,
\begin{equation}
\frac{dN}{dm}\sim m^\lambda,
\end{equation}
then the WD formation rate, and hence the DTD, will be
\begin{equation}
\frac{dN}{dt}=\frac{dN}{dm}\frac{dm}{dt}\sim t^{(1+\lambda-\delta)/\delta} .
\end{equation}
 For the commonly used value of $\delta=-2.5$ and the Salpeter
 (1955) slope of $\lambda=-2.35$, the resulting power-law index is
 $-0.46$, or roughly $-1/2$. Pritchet et al. (2008) have argued that
such a $t^{-1/2}$ form for the DTD can explain the trend of SN~Ia rate
versus specific star-formation rate of the host population in the
Supernova Legacy Survey. 

We test model power-law DTDs with 
these two particular power-law indices, $-1$, and $-1/2$, 
as well as a continuous range of indices, $s$, against the
cluster data. We assume $\Psi(t)=0$ for $t<40$~Myr, corresponding
to the lifetime of  $8{\rm M_\odot}$ stars (e.g, Girardi et al. 2000),
at the border between CC and WD formation. We start by setting the
star-formation redshift at $z_f=3$, i.e., at a cosmic time $t_f=2.1$~Gyr.

Fitting the observed cluster SN rates with a power law DTD having a free
normalization and index $s$, the best fit is obtained for
$s=-1.2$. The time-integrated SN number is $N_{\rm
  SN}/M_*=0.0044$ for a power law of this slope, which is intermediate
to  the ``minimal iron'' value we have derived above 
(Eq.~\ref{eqminiron}) and 
 the ``optimal'' value (Eq.~\ref{eqoptiron}). 
However, with the normalization thus 
unconstrained, $\chi^2$ depends weakly
on $s$, and there is a wide range of indices, $-1.6<s<2.7$, 
that give acceptable
fits to the observed rates.(We will deem as ``acceptable'' those model with 
$\chi^2_r<2$, which corresponds to a probability of $>5\%$ for 7 
d.o.f. -- 9 data points minus two free parameters, $s$ and the normalization.)
  The steeper ones among these power
laws also have sufficiently large integrals. Specifically, power laws
with $s<-0.88$ satisfy the minimal iron constraint, while shallower power laws,
e.g., $s=-0.5$,
do not.

\begin{figure}
\epsscale{1.23}
\plotone{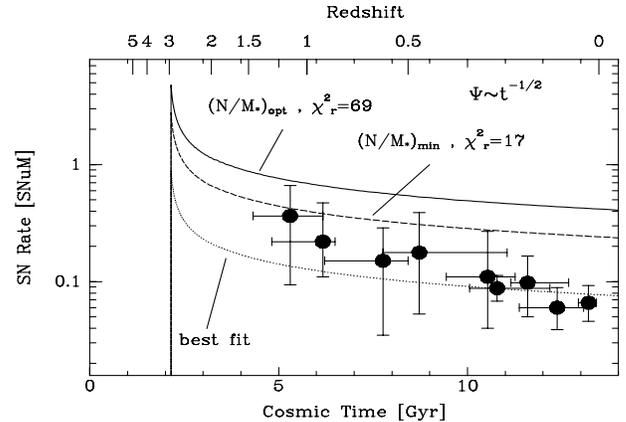}
\caption{Fits of the predictions of $\Psi\propto t^{-1/2}$ 
power-law DTDs to the observed
cluster SN~Ia rates, plotted as a function of time since the Big Bang, 
assuming instantaneous cluster star formation at $z_f=3$,
and SN~Ia events starting 40~Myr after star formation. The best-fitting
 version (dotted curve) has a time-integrated number of SNe that is 
only a minor fraction of that required to produce observed cluster 
iron-to-stellar mass ratios. Versions scaled to produce the minimal 
(dashed curve) or optimal (solid curve) integrated number of SNe give poor 
fits to the cluster rates, with the reduced $\chi^2$ values indicated.}  
\label{figplm0.5}
\end{figure}

Conversely, an $s=-0.5$ power-law
normalized to have the correct integral, when compared to the
data, grossly overpredicts the observed rates, giving 
an unacceptably high reduced chi-square, $\chi^2_r=\chi^2/{\rm d.o.f.}$,  
of 17 and 69 per degree of
freedom (d.o.f.), for the minimal and optimal normalizations,
respectively. (In the latter two fits, there are no free parameters
that are adjusted to fit the data, and hence we have 9 degrees of freedom, 
as the number of independent data points.) Figure~\ref{figplm0.5}
shows these three fits.
Thus, a $t^{-1/2}$ power law, while  it can describe well
the time dependence of the cluster rates at $0<z<1.45$, cannot simultaneously
produce the required time-integrated SN~Ia numbers. 

\begin{figure}
\epsscale{1.23}
\plotone{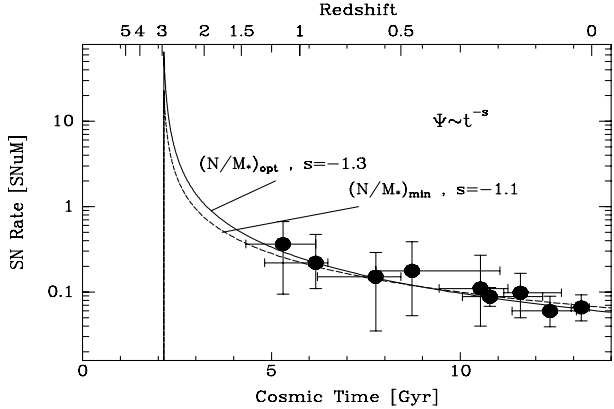}
\caption{Same as Fig~\ref{figplm0.5}, but for steeper power law DTDs, 
$\Psi\propto t^{-s}$, 
with $s=-1.1$ and the minimum-iron normalization (dashed curve), and
$s=-1.3$ plus the optimal-iron normalization (solid curve). In both cases,
there is a good fit to the cluster SN rate data.}
\label{figplm1}
\end{figure}

 If we force the
minimal-iron constraint, the best-fit power law is $t^{-1.1}$, with 
an acceptable $\chi^2_r=0.11$, as shown in Fig.~\ref{figplm1}.
The low $\chi^2$ value indicates that the errors on at least some of the
rates have been conservatively overestimated. 
 The combined constraints of
the SN rate data and the minimal iron abundances limit the acceptable values
of the power law index to $s=-1.10_{-0.22}^{+0.28}$.
 For 
the optimal iron constraint, the acceptable range of indices 
is  $s=-1.28^{+0.25}_{-0.18}$, with 
$\chi^2_r=0.10$ for the best fit. Smaller, more realistic, rate uncertainties
would reduce these allowed ranges.

 We thus see that a simple
parametrized model DTD of the form $t^{-1}$, or slightly steeper,
can provide a good fit to the cluster SN~Ia rates while
simultaneously providing a sufficient time-integrated number of SNe~Ia
to satisfy the iron-based constraints. 
   
We note that these conclusions depend weakly on the chosen epoch 
of cluster star formation, $t_f$. For a given power-law slope,
acceptable fits are obtained for $t_f$ in the range 1.2 to 2.3~Gyr,
corresponding to $z_f=4.8$ to 2.8, with a best fit that has a constant
minimum $\chi^2$ in the range $z_f=3.2-3.7$.
Varying $t_f$ cannot salvage power-law DTDs such as $t^{-1/2}$, 
that are strongly ruled out because of the shallowness of their slope. 
However, moving $t_f$ back can slightly increase the allowed range
of slopes. For example, for the optimal iron value, the maximal
acceptable power-law slope can be raised from $-1.11$ to $-1.08$ by 
shifting $z_f$ back to 3.6.  
 
It is arguable that, instead of 
 a single, $\sim t^{-1}$ power law, motivated by binary mergers, 
with this power law
  extending back to delays as short as
 40~Myr, there could be a ``bottleneck'' in the supply of progenitor
systems below some delay. Such a bottleneck could be 
 due to the birth rate of WDs, which 
behaves as $\sim t^{-1/2}$. One possible result would then be a
 broken-power-law DTD, with $\Psi\propto t^{-1/2}$ up to some
 characteristic time, $t_c$, and $\Psi\propto t^{-1}$ thereafter.
A possible value could be  $t_c\approx 400$~Myr, corresponding to the 
lifetimes of $3M_\odot$ stars. If that were the lowest initial mass of
 stars that can produce the WD primary in a DD SN~Ia progenitor, then
beyond $t_c$ the supply of new systems would go to zero, and the 
SN~Ia rate would be dictated by the merger rate.

We have therefore attempted to fit the data with such a $t^{-1/2}$, $t^{-1}$
broken power law. With the DTD normalization fixed to produce the required
mimimum iron integrated SN numbers, 
this model gives an acceptable $\chi^2_r=1.2$,
though this is considerably worse than the single power-law fit, and could
become unacceptable with more realistic errors.
We test the optimal iron value with a 
$t^{-1/2}$, $t^{-1.3}$ broken power law, that is steeper at late times
(since even $t^{-1}$ alone is already rejected, see above).
Here, too, an acceptable $\chi^2_r=0.9$ is found. 
For the minimal-iron normalization, a $t^{-1/2}$, $t^{-1.1}$ dependence
is acceptable, as long as $t_c<1.5$~Gyr. 
Such a late break time is interesting 
in the context of sub-Chandra merger models, in which the mergers
of white dwarfs of initial masses smaller than $3{\rm M_\odot}$
produce SNe~Ia (Sim et al. 2010; Van Kerkwijk et al. 2010).
    
To summarize, the cluster SN rates plus iron abundances can be fit with 
a power-law DTD, under some conditions. 
 Assuming the minimal iron mass value indicated by
cluster observations, a
range of slopes with values of $s\approx-1$, or somewhat steeper, is allowed. 
A break to 
a shallower $t^{-1/2}$ dependence at $t<t_c$
is also permitted,
provided the power-law is steep enough at longer delays,
and the break does not occur too late.
For the acceptable range of power-law DTDs, between 50\% and 85\% of SNe~Ia
explode during the first Gyr after star formation. 

\subsection{DTDs from binary population synthesis models}
\label{ss.yungel}

Over the past decade, a number of groups
(e.g., Yungelson \& Livio 2000;  Han \& Podsiadlowski 2004; 
Belczynski et al. 2008; Ruiter et al. 2009;
Bogomazov \& Tutukov 2009; Wang, Li, \& Han 2010; Mennekens et
al. 2010) have calculated DTDs using
 binary population 
synthesis (BPS), in which a large number of binaries with a
chosen 
distribution of initial conditions are followed through the stages of
stellar and binary evolution, to the point that some of them reach the
conditions for explosion as SNe~Ia. In our comparison to cluster SN
rates, we focus here, as examples, on the models
presented by Yungelson \& Livio (2000), and on the more recent ones
by Mennekens et al. (2010).
As opposed to the simple parametrized models, considered above, the
BPS models can make absolute predictions of SN rates vs. time, i.e., their 
normalizations are set. Therefore, in our comparisons of these models
to the observed rates and to the integrated iron mass in clusters, we 
first consider the ``raw'' predictions of the models, without any
scalings, and then proceed to test scaled versions.

Yungelson \& Livio (2000) studied 
four different
evolutionary paths to a  SN~Ia: a DD model; an SD model with 
accretion of He from a giant companion and
detonation at sub-Chandrasekhar mass, through an edge-lit detonation
caused by ignition of the He layer (He-ELD); and SD models with accretion from
a sub-giant companion and detonation at the Chandrasekhar mass (SG-Ch), or
through an edge-lit detonation (SG-ELD). The DTDs for these
different paths can be seen in their fig. 2. We have scaled up these
DTDs by a small factor of 1.05, to convert from the IMF assumed by Yungelson \& Livio (2000) to
our adopted diet Salpeter IMF. Their assumed IMF is a broken power
law, of index $\lambda=-2.5$ from $0.3 {\rm M_\odot}$ to $100 {\rm M_\odot}$, 
and with $\lambda=0$ from $0.08 {\rm M_\odot}$ to $0.3 {\rm M_\odot}$ (L. Yungelson,
private communication). 
(For a fully self-consistent
comparison of the shapes of the DTDs, their models would need to be 
re-calculated with our adopted IMF, but it is unlikely that the
different IMF slopes, $-2.5$ and $-2.35$, over the limited range of
masses that contribute to SN~Ia progenitors, would lead to major
changes in the DTD).
 Two of the models, He-ELD and
 SG-Ch, predict no SNe beyond $1.5-2$~Gyr after star formation, where 
all the measured cluster SN rates are. These models are obviously 
inconsistent with the observed rates, although they could play a role
in a multi-component DTD scenario (see \S\ref{ssdoublecomp}, below). 
The  DD and SG-ELD models, on the other hand, do
 predict SN events on long time scales, and can be tested. 

\begin{figure}
\epsscale{1.23}
\plotone{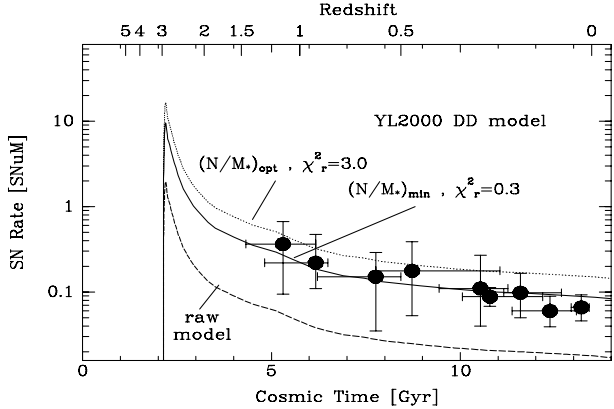}
\caption{Comparison of the observed cluster SN rates to the predictions
of the DD model from the BPS simulations of Yungelson \& Livio (2000).
The ``raw'' prediction of the model (dashed curve), 
without any scaling, underpredict
both the observed rates and the required integrated number by a factor of 
5. A version of the model (solid line) scaled to produce the minimal-iron
normalization matches well the observed rates. When scaled to 
the optimal-iron value, however, it overpredicts the low-redshift 
cluster rates.}  
\label{figyungdd}
\end{figure}
Figure~\ref{figyungdd} compares the predictions of 
the DD model of Yungelson \& Livio (2000) to the observations.
The ``raw'' version of this model, without any
re-scaling of the DTD (beyond the adjustment above for conversion
between the IMFs), underpredicts the observed cluster SN rates, with 
a poor $\chi_r^2=2.9$. Its integrated number of SNe per formed stellar
mass is $N_{\rm
  SN}/M_*=0.0007$, just 1/5 of the minimal-iron value, 
and only 12\% of the optimal value. 
This confirms previous assertions by Maoz (2008), Ruiter et
al. (2008), and Mennekens et al. (2010), that BPS models underpredict
observed SN rates by at a factor of at least a few, and likely by more.  
As seen in Fig.~\ref{figyungdd},
if we scale up the model by a multiplicative factor of 5, 
so as to integrate to the 
minimal-iron value, we obtain an acceptable $\chi_r^2=0.3$ for
the rates. Although BPS models have many free parameters, it is not clear
that such a level of scaling-up of the model could be achieved easily.
If we force the DTD integral to the optimal-iron value, the 
predicted rates are too high, giving an unacceptable $\chi_r^2=3.0$.  

\begin{figure}
\epsscale{1.23}
\plotone{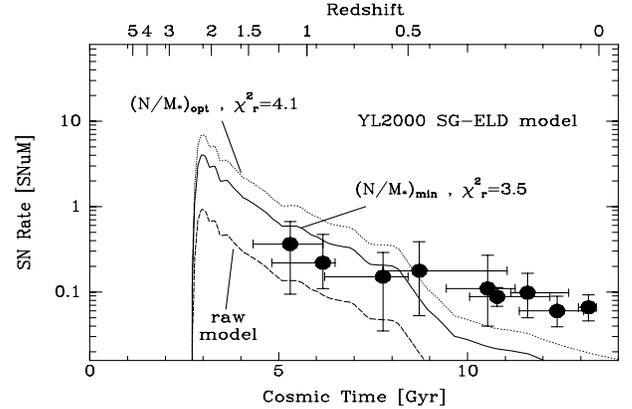}
\caption{Same as Fig.~\ref{figyungdd}, but for the SD model 
of Yungelson \& Livio (2000) involving a sub-giant donor and an edge-lit 
detonation. The cluster epoch of star formation, as before, is
assumed to be $z_f=3$, but in this model the first SN~Ia events are delayed
by $\sim 800$~Myr. Because of the steep decline of this model DTD,
neither the raw models, nor those scaled to the normalizations required
by iron abundances, match the rate observations.   
}  
\label{figyungsgeld}
\end{figure}

Proceeding to the SG-ELD model of Yungelson \& Livio (2000), 
this model (Fig.~\ref{figyungsgeld}) always gives a poor fit to
the SN rates, whether in its raw form (which again produces only
23\% of the minimum iron value), or scaled to satisfy the
iron constraints, or even if scaled arbitrarily. The SG-ELD model
begins making SNe~Ia only about 800~Myr after star formation, and
thus misses the opportunity of producing the bulk of the iron mass
during that time. Furthermore, this model then predicts SN rates that fall too
steeply with time. These problems cannot be alleviated by a change in
the star-formation epoch, $t_f$, which has little effect on the fits,
as was the case for parametrized power-law DTDs.

We turn now to the BPS models of Mennekens et al. (2010), who have
examined both DD and SD models. They have assumed a Kroupa et
al. (1993) IMF, consisting of a three-part broken power law, with
index  $\lambda=-1.3$ from $0.08 {\rm M_\odot}$ to $0.5{\rm M_\odot}$,
 $\lambda=-2.2$ from $0.5 {\rm M_\odot}$ to $1{\rm M_\odot}$, and
$\lambda=-2.7$ from $1 {\rm M_\odot}$ to $100 {\rm M_\odot}$. 
The mass ratio between this IMF and the diet-Salpeter IMF is 0.92, 
and we therefore scale down the Mennekens et al. (2010) DTDs by this factor.

In their DD models that we examine, Mennekens et al. (2010) 
introduce the possibility of non-conservative Roche-lobe overflow,
which they parametrize with a $\beta$ parameter, the fraction of
material lost by the donor star that is accepted by the accreting star.
Values of $\beta=0, 0.8, 0.9$, and 1.0 are considered. 
The $\alpha$
parameter (Webbink 1984), 
which is the fraction of the orbital energy lost during the
common-envelope phase that is transferred to kinetic energy of the
envelope, is set to 1. 
An additional DD model is calculated with
$\beta=1$, but treating the common-envelope phase, 
instead of with the $\alpha$ model, by using
 the $\gamma$ parametrization of Nelemans \& Tout
(2005), with $\gamma=1.5$.
The $\gamma$ approach quantifies the change in angular momentum during 
the common envelope phase.
Two SD models with $\beta=1$ 
have also been calculated by Mennekens et al. (2010), one with
$\alpha=1.0$ and one with $\gamma=1.5$. 

\begin{figure}
\epsscale{1.23}
\plotone{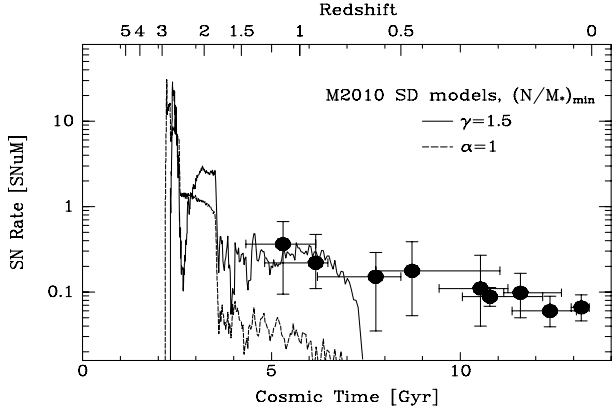}
\caption{Rate predictions based on DTDs from two SD models by 
Mennekens et al. (2010), based on two different parametrizations
of the physics of the common-envelope phase (see text). The models
have been scaled up by factors of 80 ($\gamma=1.5$ model, solid curve)
and 6.4 ($\alpha=1.0$ model, dashed curve), 
to match the minimal-iron constraint
but both fall too steeply to match the observed cluster SN rate redshift 
dependence.   
}
\label{figmensd}
\end{figure}

\begin{figure}
\epsscale{1.23}
\plotone{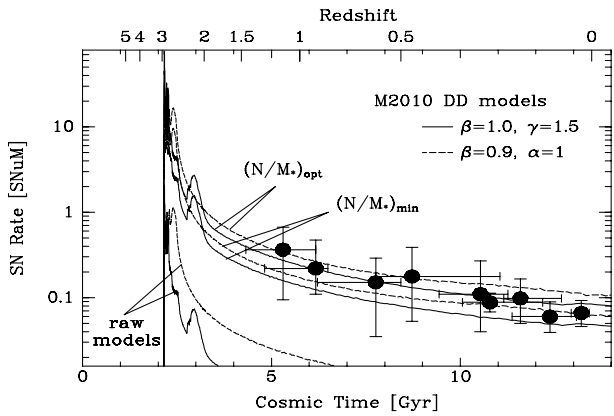}
\caption{Same as Fig.~\ref{figyungdd}, but for two of the DD models 
from the BPS calculations of Mennekens et al. (2010) that can match 
the data, if the raw DTDs are suitably scaled up according to the
optimal iron requirement, or the minimal 
requirement, as marked. The parameter $\beta$ quantifies the
degree of mass conservation during the Roche-lobe overflow phase.
}  
\label{figmendd}
\end{figure}

Fitting these models to the cluster rates, with or without the cluster 
iron constraints, we find the following. None of the raw DTDs of
Mennekens et al. (2010) produce enough time-integrated SN numbers to 
reproduce the observed iron abundances. At best, the $\beta=0.8$ and 
$\beta=0.9$ DD models produce 12\% of the minimal-iron value, and the 
SD $\alpha=1$ model makes 16\%. The
other models make only a few percent or less of the minimal number of SNe
required by cluster abundances.
If we scale up the models, forcing the
minimal-iron normalization, 
the Mennekens et al. (2010) SD models, qualitatively 
like the Yungelson \& Livio (2000) 
SD models discussed above, predict no SNe $4-5$~Gyr
after star formation, and hence cannot match the observed cluster rates.
This is illustrated in Fig.~\ref{figmensd}.
Two of the scaled Mennekens et al. (2010) DD models
give acceptable fits
to the observed cluster
SN rates: DD with $\alpha=1$ and $\beta=0.9$; and DD with $\gamma=1.5$
and $\beta=1$. 
These fits are shown in 
Fig.~\ref{figmendd}. 

To summarize our analysis of BPS models from these two teams, the emerging 
picture is that SD models always fit the observations poorly -- in
terms of both the absolute ``raw'' numbers of SNe they predict, and the 
time dependence of the cluster SN rates they predict. The raw DD models
also underpredict the absolute SN numbers, but by lower factors, of
$5-8$. If we treat the BPS models as scalable, then for some of the DD
models it is possible
to simultaneously satisfy the minimal-iron constraints and the
observed cluster SN rate dependence on redshift. 
The higher, optimal-iron, constraint can also be satisfied by two scaled
DD models among those of Mennekens et al. (2010).
 
\subsection{DTDs from analytical models}

\label{ss.greggio}
Another approach to making DTD predictions, followed by
Greggio (2005), is to calculate
analytical DTD models, based on stellar
evolution arguments and on various parametrizations of the possible 
results of the complex common-envelope phases through which SN~Ia
 progenitor
systems must pass. For each of several SN~Ia channels, she calculated the DTDs
that emerge when varying the values for the parameters describing the
 initial conditions, and the mass and separation distributions and 
limits of the systems that eventually explode.  
Mennekens et al. (2010) have criticized this approach in that it
overlooks the changes in stellar-evolution timescales that occur as 
a result of the ``rejuvenation'' due to mass transfer between stars.
On the other hand, 
an advantage of the analytic 
 approach, compared to the BPS approach, is
 that predictions for a large range of parameters can be made quickly.
This successful range can
then be investigated in more detail with actual BPS simulations.

We focus on a selection of representative
models shown in fig. 1 of Greggio et al. (2008). These include one
SD model and four DD models, computed under
different assumptions for these parameters.
The ``wide'' and ``close'' labels of the DD models refer
 to two possible
 parametric schemes used by Greggio (2005) to describe 
the WD separation distribution after the common envelope phase.
The Greggio models do not predict the absolute levels of the SN rates
(i.e., the normalization of the model DTDs), as this is another free
parameter in the models. We therefore examine only various scaled versions of
the Greggio models.

\begin{figure}
\epsscale{1.23}
\plotone{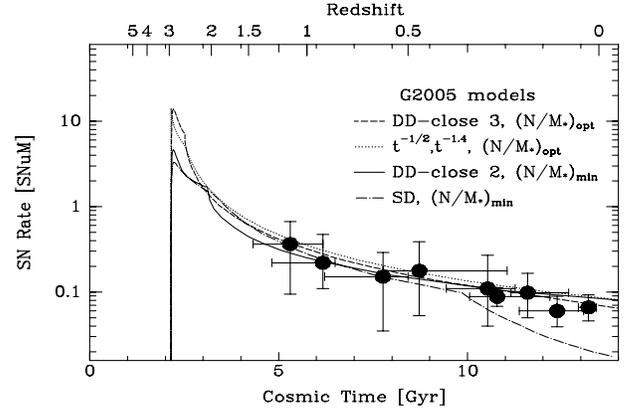}
\caption{Comparison of the observed cluster SN rates to predictions
by three of the analytic models of Greggio (2005), scaled to give
either the minimal or optimal ratio of integrated SN numbers to stellar mass. 
The SD model, like the SD models from the BPS simulations, underpredicts
the low-$z$ cluster SN rate. Two of the DD models, however, can match the 
data. Also shown, for comparison (dotted line), is the prediction
from  a simple broken-power-law DTD:  $t^{-1/2}$ at 40~Myr$<t<400$~Myr; 
$t^{-1.3}$ for $t>400$~Myr, having the minimal-iron normalization, 
which is similar to the DD-close-3 model (dashed curve).} 
\label{figgreggio}
\end{figure}

The DD-close model with a minimum secondary initial mass of $3
{\rm M_\odot}$ fits the observed SN rate redshift dependence,
while satisfying either the minimal iron constraint (with $\chi_r^2=0.52$) 
or the optimal iron constraint
(with $\chi_r^2=0.45$). Maoz et al. (2010) have already noted the 
good agreement between this Greggio (2005) model and a DTD reconstructed
from the SNe in the LOSS-SDSS subsample, and the fact that this Greggio model
is similar to a  $t^{-1/2}$,$t^{-1.3}$, broken power-law with break at
$t_c<400$~Myr. Such a broken power law was shown to 
fit well the data and the minimal-iron constraint
 in \S\ref{sspowerlawdtd}, above. The resemblance
is seen in Fig.~\ref{figgreggio}. The DD-close model with a minimum 
secondary mass of $2{\rm M_\odot}$, also shown in  Fig.~\ref{figgreggio}, can
fit the rates as well, but only with the minimal-iron constraint,
due to its shallower slopes at both early and late times.
The two DD-wide models of Greggio (2005) we test, with minimal masses
of $2{\rm M_\odot}$ and $2.5{\rm M_\odot}$, are both too shallow to fit the rates
and either of the  normalizations, and give $\chi^2_r>3.6$.  
 
Finally, the SD model of Greggio (2005), while having a DTD that is
more extended in time
than the SD models from BPS that we have examined above, suffers
from similar problems, particularly a steep drop in rates beyond $\sim 8$~Gyr.
The minimal-iron normalized SD model, shown in  Fig.~\ref{figgreggio},
 is formally acceptable, with 
$\chi_r^2<2$, but its low predicted rates at low redshifts are in conflict
with the measurements. (Poisson statistics would be more appropriate 
to compare the low prediction to the accurately measured, non-zero,
 rates at low redshift.)

Thus, only the two DD-close Greggio (2005) models are consistent with the
rates and the iron data. This is simply the result of the fact that DD
models generically produce power-law DTDs, and the
power-laws required to fit simultaneously
the observed rates and the normalizations 
 need to have indices of $\approx -1$ or steeper. Such indices can 
be obtained by emerging from the common-envelope phase with a steep power-law 
separation distribution, i.e., with relatively more close pairs,
as discussed in \S\ref{sspowerlawdtd} and seen in Eq.~\ref{DDdependence}.
Naturally, we have examined a limited range of the Greggio (2005)
models, and it would be interesting to see if there are others
(e.g. Greggio 2010) that do
fit the data, and what are their parameters.

\section {Comparison to DTD predictions -- non-instantaneous 
cluster star-formation histories}
\label{ss.noninst}
So far, we have assumed the cluster SFH to be a single, instantaneous 
burst at $z_f$. We now test whether relaxing this assumption can improve
the fit of any of the models to the data, in terms of reproducing
both the time dependence of the SN~Ia rate and the normalization,
as required by the iron abundance. Furthermore, a non-instantaneous
burst may be a more realistic description of cluster SFH. 
The predicted SN rate will now be a convolution of the SFH, $S(t)$, with
the DTD, after correction for
stellar mass loss,
\begin{equation}
R_{Ia}(t)\varpropto\int_{0}^{t}S(t-\tau)\frac{\Psi(\tau)}{m(\tau)}d\tau.
\end{equation}

We 
consider a single, but non-instantaneous, burst of
 star formation in galaxy clusters in the form of an 
exponentially decaying SFH, $S(t)\propto {\rm exp}[-(t-t_f)/\tau_{\rm SF}]$,
starting at time $t_f$. More complex SFHs are considered
in \S\ref{ss.compositesfh}, below.
We have re-fit the cluster rates with the iron-mass normalizations using
all the DTD models discussed in
\S\ref{sscomparisontopred}, above, but convolved with this exponential SFH, 
with values of $\tau_{\rm SF}=0.5,1,2,3,4$~Gyr. The effect on the SN rate
of the convolution
between the DTD and any temporally extended star formation is always to
transfer  some fraction 
of the SN events to later times. Since all the DTDs that 
we consider 
peak at short delays, this means that the SN rate rises more slowly at early
times than in the instantaneous burst approximation. This, in turn, means
that a smaller fraction
 of the time-integrated SN number, dictated by the iron abundances,
can be produced at early times. Raising the model normalization, such that
the iron constraints are met, then results in a poor fit to the SN rate data,
compared to the instantaneous case. To try to mitigate this effect, we
shift back by 1~Gyr the time of initial star formation
 to $t_f=1.1$~Gyr, corresponding
to $z_f=5$. Nonetheless, in every case, the fit with an extended SFH
is worse than in the case of an
instantaneous burst.

\begin{figure}
\epsscale{1.23}
\plotone{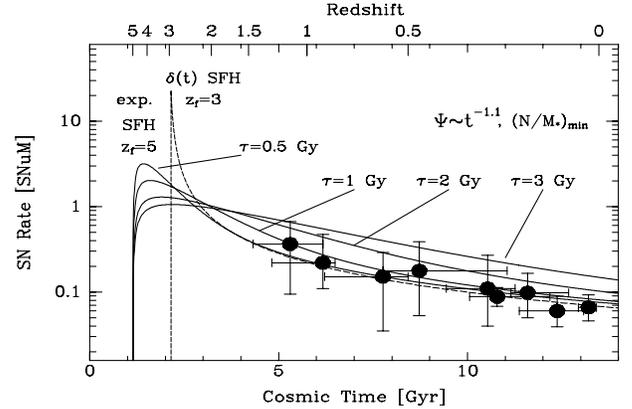}
\caption{Comparison of the data to 
predictions of a  $t^{-1.1}$ power-law DTD, convolved
with an exponential SFH starting at $z_f=5$, and several characteristic 
exponential times
$\tau_{sf}$. The prediction for the same DTD, but following an 
instantaneous burst at $z_f=3$,
shown before in Fig.~\ref{figplm1}, is plotted for comparison (dotted curve).
 All curves
are normalized to satisfy the minimal-iron constraint.
Such extended SFHs degrade the agreement with observations, and 
$\tau_{sf}<2.7$~Gyr is required to avoid significantly 
overpredicting the rates. 
}
\label{figplsfhexp}
\end{figure}

\begin{figure}
\epsscale{1.23}
\plotone{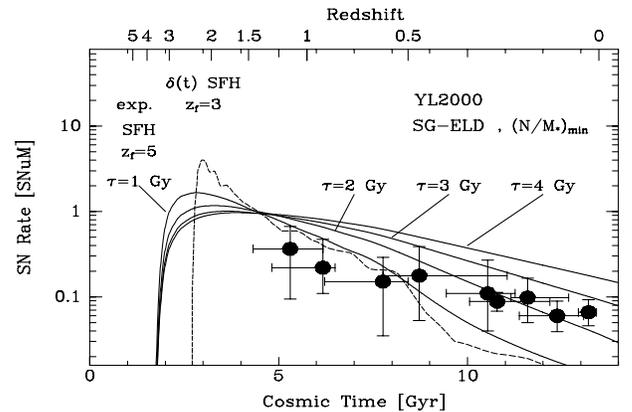}
\caption{Same as Fig.~\ref{figplsfhexp}, but for the Yungelson \& Livio (2000)
SG-ELD model, previously shown in Fig.~\ref{figyungsgeld}. The convolution
of this DTD with an extended exponential SFH cannot improve the fit
to the SN rates, if the iron constraints on the normalization are 
to be simultaneously satisfied.
}
\label{figyungsfhexp}
\end{figure}

These results are illustrated with two examples. Figure~\ref{figplsfhexp}
shows the predictions of a minimal-iron-normalized $t^{-1.1}$ power-law DTD,
with the exponential SFH starting at $z_f=5$, and several characteristic 
exponential times
$\tau_{sf}$. Also shown, for comparison, is the $z_f=3$ instantaneous burst,
shown before in Fig~\ref{figplm1}. At late times, all the predictions
have similar slopes, while at progressively shorter times, the rate
dependence is shallower for the more extended SFHs. This leads to 
progressively greater overprediction of the observed rates. Formally, 
$\tau_{sf}>2.7$~Gyr is ruled out, based on $\chi_r^2>2$. Conversely, if we
ignore the iron constraints and find the best-fit normalization, then the 
integrated number of SNe~Ia is significantly lower than required by the
minimal constraint (e.g., by a factor 2, for $\tau_{sf}=3$~Gyr).
In Fig.~\ref{figyungsfhexp}, we show the same exercise for the SG-ELD model
of Yungelson \& Livio (2000). It might have been hoped that the convolution
with an extended SFH could moderate the steep fall at late delays of this
DTD, and thus provide a better fit to the observed 
weak time dependence of the rates.
Although the steep fall, predicted in the instantaneous burst case, is
indeed moderated, the smoothing at short delays lowers the contribution
from early times to the integrated SN numbers, forcing a higher overall
normalization and a poor fit.
  
To summarize this section, the moderate slope of the
 observed SN rate redshift dependence
at $0<z<1.4$.,
combined with the large time-integrated number of SNe~Ia indicated by 
the iron abundance, together call for a DTD that is sharply peaked at 
short delays, but with a low tail out to long delays. Convolution of any
of the few single-DTD models that satisfy these constraints with any
simple SFH that is extended 
on timescales $\gtrsim 1$~Gyr only degrades the fits.

\section{Comparison to DTD predictions -- two-component models}
\label{ssdoublecomp}
We now examine to what degree the challenges of reproducing the observations 
can be overcome by the
addition of free parameters that is implicit in
the combination of multiple components
 -- either two DTDs (as could be expected from the co-existence of two
 distinct physical SN~Ia channels, e.g., DD and SD), or two
components of cluster SFH, as opposed to the single bursts assumed so far. 
  
\subsection{Double DTD models}
\label{ssdoubledtd}
The idea of two co-existing SN~Ia channels, prompt and delayed,
emerged several years ago from the observation of, on the one hand, 
a proportionality between star-formation rate (SFR) 
and SN~Ia rate per unit stellar mass
in star forming galaxies,
and on the other hand, a non-zero SN~Ia rate in early-type galaxies
with no current star formation 
(Mannucci et al. 2005, 2006; Scannapieco \& Bildsten
2005; Sullivan et al. 2006). 
As noted in \S\ref{ss.intro},
this observation does not necessarily imply the existence of two 
separate physical channels. Instead, it could be the manifestation of a
DTD from a single channel but with delays
 spread over a wide range of timescales. Indeed, our analysis, above,
shows that single parametrized DTDs of power-law form, with indices somewhat 
steeper  than $-1$ can match the observational constraints (as can some
DD models that produce DTDs of this type, if they are suitably scaled up). 
However, some of the other theoretical DTDs, that individually are incompatible
with the data because they predict no delayed SNe,  can be ``saved''
by incorporating them into a two-channel picture.

\begin{figure}
\epsscale{1.23}
\plotone{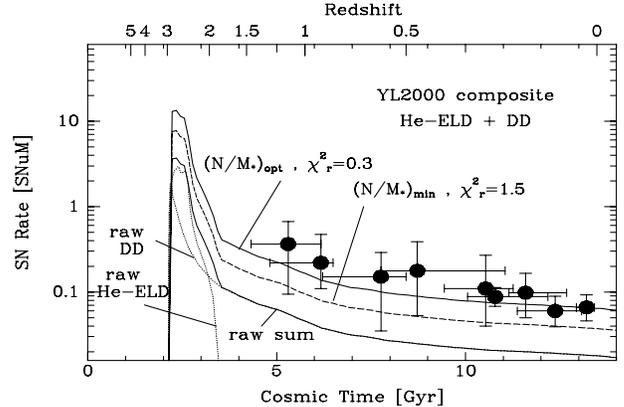}
\caption{Predicted SN rate dependence for a composite DTD model, combining
the He-ELD and DD models of Yungelson \& Livio (2000). Dotted curves show
the raw predictions of these models, and lower solid line is their sum.
The raw sum underpredicts the low-redshift rates by a factor of 5 (as 
was the case for this  DD model alone, Fig.~\ref{figyungdd}), and the
integrated SN number by 2.1 (minimal iron) to 3.6 (optimal iron). Scaled-up
versions of these models do give a satisfactory fit to all the observations,
as indicated.}
\label{figheeldpdd}
\end{figure}

For example,
two of the models of Yungelson \& Livio (2000),  He-ELD and the SG-Ch,
predict no SNe at long time delays. Figure~\ref{figheeldpdd}
shows the predicted cluster rates from the combination 
of the He-ELD model and the DD  model, discussed in \S~\ref{ss.yungel}.
The raw sum of these two models, as before, produces a too-small fraction
(less than half) of the total number of SNe~Ia indicated by the minimal iron 
constraint, and underpredicts all the SN rates below 
$z<0.5$ (by a factor of 5).
However, as seen in Fig.~\ref{figheeldpdd},
scaling up this composite DTD by factors of 2.1 or 3.6 can solve
these problems for  the minimal and optical iron cases, respectively. 
In this example, 60\% of the SNe (and the iron) are from the prompt He-ELD
SD component, and the rest from the DD component.

Similar combinations of two components can work using the other DTDs
we have considered, as long as one of the components is 
a DD model (or a similar power law) that can provide the SNe with long delays.
In addition to the choice of components, one can choose the relative scaling 
between them, providing a further adjustable parameter.
The current data
of course cannot discriminate between  the various possible prompt components
as they make their contributions beyond the redshifts at
which rate measurements exist.
Our experiments at combining different DTD are obviously not
exhaustive, but the emerging picture is nonetheless clear. The
observed SN rates and the iron mass in clusters can be explained
simultaneously by combining ``prompt'' and ``delayed'' DTDs.
The SNe~Ia from the prompt component produce the majority of the 
iron mass in clusters. The SNe from the delayed component produce
only a fraction of the metals, and they are the events detected by
current SN rate measurements, with their weak time dependence.

\subsection{Composite star formation histories}
\label{ss.compositesfh}
\begin{figure}
\epsscale{1.23}
\plotone{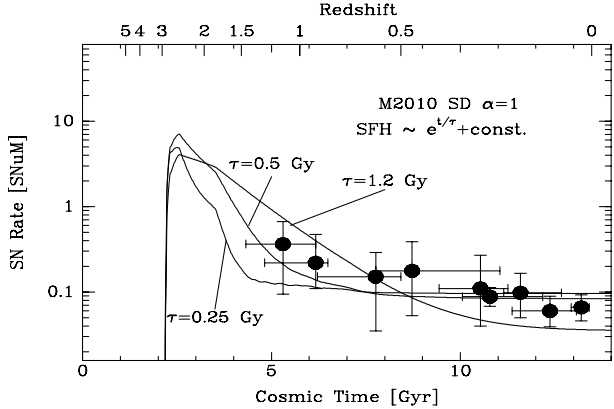}
\caption{Same as Fig.~\ref{figheeldpdd}, but with a single, prompt, DTD,
and a composite SFR -- an exponentially decaying burst at $z_f=3$, with
characteristic times as labeled, and a constant ``DC'' component of star 
formation, with a ratio between the components of 200:1 at $z=3$.
 The DTD is the $\alpha=1$ SD model of Mennekens et al. (2010),
previously shown if Fig.~\ref{figmensd}. Normalizations are minimal-iron 
for the short-timescale burst and optimal-iron for the two longer-timescale
bursts. As seen in the figure, such composite SFHs can match the rate
observations and the iron constraints with a single, prompt, DTD.
However, the constant and high level of star formation down to low 
redshifts is at odds with other observations of cluster galaxies.
 }  
\label{figmensfexpplusconst}
\end{figure}

A final scenario we examine is that of a single DTD, but with 
a composite cluster SFH, consisting of 
a short starburst, beginning at $z_f$, combined with a more
extended SFR. With
the additional free parameters introduced in this scheme, 
it is easy to find combinations of DTDs with such composite SFHs that 
reproduce the SN~Ia rate versus time, 
while simultaneously 
providing a sufficient time-integrated number of SNe to produce 
the observed ratio of iron mass to stellar mass.
Figure~\ref{figmensfexpplusconst} shows an example. Here, we have used
the $\alpha=1$, $\beta=1$ SD model,
 discussed previously in \S\ref{ss.yungel}, from the 
BPS calculations of Mennekens et al. (2010). We recall that this DTD
produces only SNe~Ia with short delays. In the context of the 
present exercise, this is 
desirable, as the observed SN rate dependence in generated by the
SFH rather than by the DTD. In this example, for the two SFH components 
we take the sum, $S(t)=S_1(t)+S_2(t)$, of an exponential SFR, 
$S_1(t)\propto {\rm exp}[-(t-t_f)/\tau_{\rm SF}]$ as in \S\ref{ss.noninst},
and a constant SFR, $S_2(t)={\rm const}$. 
The relative levels of the two components are adjusted
to fit the iron-based normalization constraints. For this choice of functions,
$\tau_{\rm SF}<4$~Gyr is required. A more prolonged exponential component 
either overpredicts the high-$z$ rates, or forces the constant component
to overpredict  the low-$z$ rates.

This scenario can, in principle, explain the presently discussed data 
in terms of ``prompt'' DTDs such as those from SD models,
combined with residual
star formation. However, it is
at odds with many other observations of clusters.
Star-formation activity in clusters
avoids the cluster cores ($\lesssim 1$~Mpc), 
and increases progressively with radius
(e.g., Hansen et al 2008;
Porter et al. 2008; Bai et al. 2009; Saintonge et al. 2008; Loh et
al. 2008). In contrast, all of the 
cluster-SN candidate hosts in Sharon et al. (2010) 
are found at
projected distances $< 0.7$ Mpc from the brightest cluster galaxy.
In the sample of  Barbary et al. (2010) all are within $<0.8$~Mpc,
or even $<0.5$~Mpc if excluding the outermost one
\footnote{It would be
useful  to study the dependence of cluster SN
rates on radial distance from cluster cores. Such an analysis may be
feasible, however, only with a larger sample of cluster SNe.}. 
Furthermore, the
SNe from the various
surveys we analyze are almost always  
found in early-type, red-sequence (and hence
apparently quiescent), galaxies. For example, in the sample of 
Gal-Yam et al. (2008), for 4 out of the 5 cluster SNe~Ia that have host
galaxies, the galaxies are early-types, based on colors and spectra.
In the sample of Sharon et al. (2010), 6 out of the 7 most likely
cluster SN~Ia hosts are red-sequence galaxies. In Barbary et
al. (2010), this figure is 8 out of 9. Thus, the cluster SNe~Ia, on
which are based the rates that we analyze, are neither at the locations nor 
in the types of galaxies where star-formation in clusters is actually
observed to occur.
  
Quantitatively,
in the above version of the burst+constant SFH scenario, as shown in 
Fig.~\ref{figmensfexpplusconst},  
for, e.g., $\tau_{\rm SF}=0.5$,
$\sim 93\%$ of the present-day cluster stellar mass is formed before
$z=1$, and the remaining 7\%  is
 from the constant SFR between $z=1$ and 0.
A typical monitored stellar mass in the clusters we consider
is  $\sim10^{13}$ M$_{\odot}$ (see \S~\ref{ssgastostar}) or,
correcting to formed mass before stellar evolution mass losses, 
$\sim 2\times 10^{13}$ M$_{\odot}$. With 
a lookback time to $z=1$ of 8~Gyr, the 7\% fraction
implies a constant SFR between $z=0$ and 1 of  
$\sim$175 M$_{\odot}$ yr$^{-1}$, just in the central regions of
clusters.

By comparison,  several recent studies (e.g., Krick et al. 2009,
 Bai et al. 2009; Koyama et al. 2010) find SFRs,
integrated over the galaxies within a
cluster, between $\sim 10$ and a few hundred M$_{\odot}$
yr$^{-1}$. Some clusters thus do have the high 
integrated SFRs required in the scenario of a
composite SFH + prompt DTD (ignoring for the moment that this
SFR is not in the central, quiescent galaxies that are seen to host
the SNe, but rather in the outer regions, e.g., Haines et al. 2009). 
However, the observed cluster SFR, normalized by
cluster mass, rises steeply with
redshift, as $(1+z)^{p}$, with $p=5.3\pm 1.2$ (Bai et al. 2009),
$p=5.7^{+2.1}_{-1.8}$ (Haines et al. 2009), or $p\approx 6$ (Koyama et
al. 2010). This 
corresponds to one or two orders of magnitude increase
in SFR over the $0<z<1$ range, and is in contrast to the roughly
constant mass-normalized SFR that is required in this range, 
in the composite SFH scenario, to reproduce
the cluster SN rates. 

 We also note that, if the DTD were universal in environment and in time,
and the
SFH + DTD scenario were true,  then it 
would apply to early-type galaxies in general.
SNe~Ia in all early types, whether in the field or
in clusters, and at low or high redshifts, would be the prompt outcome
of low levels of recent but unseen star-formation in these galaxies.
However,
a study by Foerster \& Schawinski (2008) of the early-type
hosts of SNe~Ia argues against this possibility. Furthermore, the DTD
reconstruction in nearby galaxies by Maoz et
al. (2010) demonstrates a $4\sigma$ detection
 of a delayed SN~Ia component, with 
delays of $2.4<\tau<13$~Gyr. 

We therefore conclude that the scenario of a burst+constant SFH,
 combined with a
prompt single-component DTD, is probably not a viable model.
 While it is capable of explaining the cluster SN rate data and 
the observed iron to stellar mass ratio,
it is incompatible with other cluster observations. Those observations
indicate that, although about 10\% of the stellar mass of clusters
 formed at $z<1$, this activity did not occur in the central,
 quiescent, galaxies hosting the SNe found by current surveys, 
and that the SFR falls steeply with cosmic time, in
 contrast to the flat SFH required to reproduce, under this scenario,
 the flat observed SN rate.

\section{Conclusions}

Measurements of 
galaxy-cluster SN~Ia rates as a function of cosmic time provide a 
particularly direct avenue to obtain the DTD of SNe~Ia, with its 
implications for progenitor models and cosmic history. Such
measurements can constrain both the functional form and the normalization
of the DTD.
We have combined recently completed measurements of cluster rates
between $z=0$ and $z=1.45$
with the latest results on the iron mass content of clusters, to recover
the SN~Ia DTD, and to test a variety of model DTDs,
from purely mathematical
parametrized forms, to DTDs obtained from more detailed physical
considerations.

Our main results are as follows.\\
1. Assuming that the bulk of the stars in clusters formed with a normal IMF 
in a brief burst
at $z_f\approx 3$, as indicated by optical spectroscopy of cluster galaxies,
the SN~Ia DTD can be directly recovered from the observations. The resulting
DTD peaks at the shortest delays probed, $0<t<2.2$~Gyr, decreases
steeply at longer delays, and extends out to 11~Gyr. The recovered DTD
agrees remarkably well, both in shape and in absolute normalization, with 
DTDs recently recovered using different techniques, 
in different
environments: in field ellipticals (Totani et al. 2008); in nearby galaxies
(Maoz et al. 2010); and in the Magellanic Clouds (Maoz \& Badenes 2010).
The current derivation is complementary to previous ones in that
it recovers the DTD over its full time range, with good time resolution
at delays $\gtrsim 3$~Gyr. The emerging picture is of a universal 
DTD that is not strongly dependent on environment or cosmic time, and can
be well represented by a single power law, of index $s \sim -1$.\\
2. Comparing the data to theoretical or phenomenological 
DTDs in a forward-modeling process, 
the best parametric description of the DTD is a power law,
\begin{equation}
\Psi(t)\approx 0.01~{\rm SN~yr}^{-1}(10^{10}~{\rm M_\odot})^{-1}
\left(\frac{t}{\rm 1~Gyr}\right)^s
\end{equation}
 with index 
$s=-1.1\pm 0.2$ or $s=-1.3\pm 0.2$, depending on whether we adopt a 
minimal-iron constraint or an optimal-iron constraint, respectively, based
on cluster data. Single-component DTDs consisting of shallower power
laws, such $\Psi\propto t^{-1/2}$, cannot 
simultaneously match the observed SN rates and the integrated SN~Ia numbers 
dictated by the iron-to-stellar mass ratio.\\ 
3. Physical DD models from the BPS simulations we have examined
can match the observations, provided they are scaled up in numbers by factors
of $5-8$. On the other hand, SD models, on their own, fail because they do 
not produce SNe at late delays, as implied by the data. Our results thus
provide strong support for the double-degenerate SN~Ia progenitor scenario.\\
4. The above conclusions are insensitive to the exact epoch of star formation 
in clusters, in the range $z_f=2-5$. The conclusions also do not change
if one replaces the instantaneous starburst with an exponentially decaying
SFH, except that the model fits deteriorate with increasing star-formation
timescale, and so the range of acceptable models shrinks.\\
5. Multi-component models, that either combine freely scaled versions of 
two DTDs (e.g. SD and DD),
or two SFH models (such as a burst plus a constant) with a prompt DTD, 
have enough freedom 
to permit many combinations that can match the existing data. However,
apart from the loss of simplicity involved, such combinations face
other problems. For the composite DTD models, it remains to be seen
if the required combinations and scalings are physically plausible. For the composite
SFHs, a high SFR in the central regions of 
clusters, and which remains constant between $0<z<1$, is required, 
in conflict with observations of clusters at these redshifts.\\
6. All of the successful DTD models produce just a  fraction of the
time-integrated numbers of SNe~Ia at redshifts below 1.4. A clear
prediction of these models is, therefore, that SN surveys of clusters
 or proto-clusters at even higher redshifts, approaching the stellar
 formation redshifts of the clusters, should reveal a sharply rising 
rate of SNe~Ia. Alternatively, observation of a non-rising SN~Ia rate
would raise the possibility 
that the bulk of metals in clusters was produced by CC SNe
from a top-heavy IMF that exploded even earlier
in cluster environments.

\acknowledgements
We thank K. Barbary, M. Graham, F. Mannucci,
N. Mennekens, E. Ofek, D. Poznanski,
C. Pritchet, L. Yungelson, D. Zaritsky, and the anonymous referee for
providing useful input. 
D.M. acknowledges support by a grant from the Israel Science Foundation.
A.G. acknowledges support by the
grant 07AST-F9 from the Ministry of Science, Culture and
Sport, Israel, and the Ministry of Research, France. A.G.
is also supported by the Israel Science Foundation,
the EU FP7 Marie Curie program via an IRG fellowship, the 
Benoziyo Center for Astrophysics, Weizmann-UK, 
and the Peter and Patricia Gruber Awards.
K.S. thanks the Benoziyo Center for Astrophysics for its hospitality
in the course of this work.
This work was supported by grants GO-10493 and GO-10793 
from the Space Telescope Science
  Institute, which is operated by the Association of Universities for
  Research in Astronomy, Inc., under NASA contract NAS 5-26555.


\begin{thebibliography}{}
\bibitem[Anders
\& Grevesse(1989)]{1989GeCoA..53..197A} Anders, E., \& Grevesse, N.\ 1989, \gca, 53, 197
\bibitem[Anderson et al.(2009)]{2009ApJ...698..317A} Anderson, M.~E.,
Bregman, J.~N., Butler, S.~C., \& Mullis, C.~R.\ 2009, \apj, 698, 317

\bibitem[Andreon et al.(2008)]{2008MNRAS.385..979A} Andreon, S., Puddu, E.,
de Propris, R., \& Cuillandre, J.-C.\ 2008, \mnras, 385, 979

\bibitem[Andreon(2010)]{2010arXiv1004.2785A} Andreon, S.\ 2010,
arXiv:1004.2785
\bibitem[Aubourg et
al.(2008)]{2008A&A...492..631A} Aubourg, {\'E}., Tojeiro, R., Jimenez, R., Heavens, A., Strauss, M

\bibitem[Badenes et al.(2010)]{2010arXiv1003.3030B} Badenes, C., Maoz, D.,
\& Draine, B.\ 2010, arXiv:1003.3030, MNRAS, in press

\bibitem[Bai et al.(2009)]{2009ApJ...693.1840B} Bai, L., Rieke, G.~H.,
Rieke, M.~J., Christlein, D., \& Zabludoff, A.~I.\ 2009, \apj, 693, 1840

\bibitem[Balestra et al.(2007)]{2007A&A...462..429B} Balestra, I., Tozzi, P., Ettori, S., Rosati, P., Borgani, S., Mainieri, V., Norman, C., \& Viola, M.\ 2007, \aap, 462, 429 
\bibitem[barbary]{brb}Barbary, K., et al. 2010, ApJ, submitted
\bibitem[Barris \& Tonry(2006)]	{2006ApJ...637..427B} Barris, B.~J., \& Tonry, J.~L.\ 2006, \apj, 637, 427

\bibitem[Belczynski et al.(2008)]{2008ApJS..174..223B} Belczynski, K.,
Kalogera, V., Rasio, F.~A., Taam, R.~E., Zezas, A., Bulik, T., Maccarone,
T.~J., \& Ivanova, N.\ 2008, \apjs, 174, 223

\bibitem[Bell et al.(2003)]	{2003ApJS..149..289B} Bell, E.~F., McIntosh, D.~H., Katz, N., \& Weinberg, M.~D.\ 2003, \apjs, 149, 289

\bibitem[Boehringer et al.(2005)]{2005Msngr.120...33B} Boehringer, H., 
Mullis, C., Rosati, P., Lamer, G., Fassbender, R., Schwope, A., 
\& Schuecker, P.\ 2005, The Messenger, 120, 33 
\bibitem[bogo]{2009ARep...53..214B}Bogomazov, A.I \& Tutukov, A.V., 2009, Astronomy Reports,
  53, 214
\bibitem[brandt]{A}Brandt, T.~D., Tojeiro, R., Aubourg, E., Heavens, A.,
  Jimenez, R., \& Strauss, M.~A.\ 2010, arXiv:1002.0848
\bibitem[Bremer et al.(2006)]{2006MNRAS.371.1427B} Bremer, M.~N., et al.\ 
2006, \mnras, 371, 1427 
\bibitem[Bruzual 
\& Charlot(2003)]{2003MNRAS.344.1000B} Bruzual, G., \& Charlot, S.\
  2003, \mnras, 344, 1000 
\bibitem[Buzzoni(2005)]{2005MNRAS.361..725B} Buzzoni, A.\ 2005, \mnras, 
361, 725 
\bibitem[Cain et al.(2008)]{2008ApJ...679..293C} Cain, B., et al.\ 2008, 
\apj, 679, 293 
\bibitem[Cappellari et al.(2006)]{2006MNRAS.366.1126C} Cappellari, M., et 
al.\ 2006, \mnras, 366, 1126 
\bibitem[Ciotti et al.(1991)]{1991ApJ...376..380C} Ciotti, L., D'Ercole, 
A., Pellegrini, S., \& Renzini, A.\ 1991, \apj, 376, 380 
\bibitem[Cooper et al.(2009)]{2009ApJ...704..687C} Cooper, M.~C., Newman, 
J.~A., \& Yan, R.\ 2009, \apj, 704, 687 
\bibitem[Daddi et 
al.(2000)]{2000A&A...362L..45D} Daddi, E., Cimatti, A., \& Renzini, A.\ 2000, \aap, 362, L45 

\bibitem[Dahlen et al.(2004)]	{2004ApJ...613..189D} Dahlen, T., et al.\ 2004, \apj, 613, 189 
\bibitem[Dahlen et al.(2008)]{2008ApJ...681..462D} Dahlen, T., Strolger, L.-G., \& Riess, A.~G.\ 2008, \apj, 681, 462 
\bibitem[Dawson et al.(2009)]{2009AJ....138.1271D} Dawson, K.~S., et al.\ 
2009, \aj, 138, 1271 
\bibitem[De Grandi 
\& Molendi(2001)]{2001ApJ...551..153D} De Grandi, S., \& Molendi, S.\ 2001, \apj, 551, 153 
\bibitem[Dilday et al.(2010)]{2010ApJ...715.1021D} Dilday, B., et al.\ 
2010, \apj, 715, 1021 

\bibitem[Ebeling et al.(2007)]{2007ApJ...661L..33E} Ebeling, H., Barrett, E., Donovan, D., Ma, C.-J., Edge, A.~C., \& van Speybroeck, L.\ 2007, \apjl, 661, L33 \bibitem[Ehlert 
\& Ulmer(2009)]{2009A&A...503...35E} Ehlert, S., \& Ulmer, M.~P.\ 2009, \aap, 503, 35 
\bibitem[Eisenhardt et al.(2008)]{2008ApJ...684..905E} Eisenhardt, 
P.~R.~M., et al.\ 2008, \apj, 684, 905 

\bibitem[Filippenko(1997)]	{1997ARA&A..35..309F} Filippenko,
  A.~V.\ 1997, \araa, 35, 309
\bibitem[fili]{B}Filippenko, A.V., et al. 2010, in preparation
\bibitem[F{\"o}rster et al.(2006)]{2006MNRAS.368.1893F} F{\"o}rster,
  F., Wolf, C., Podsiadlowski, P., \& Han, Z.\ 2006, \mnras, 368, 1893
\bibitem[F{\"o}rster 
\& Schawinski(2008)]{2008MNRAS.388L..74F} F{\"o}rster, F., \& Schawinski, K.\ 2008, \mnras, 388, L74 
\bibitem[Gal-Yam et al.(2002)]	{2002MNRAS.332...37G} Gal-Yam, A., Maoz, D., \& Sharon, K.\ 2002, \mnras,  332, 37 
\bibitem[Gal-Yam \& Maoz(2004)]	{2004MNRAS.347..942G} Gal-Yam, A., \& Maoz, D.\ 2004, \mnras, 347, 942
\bibitem[Gal-Yam et al.(2008)]{2008ApJ...680..550G} Gal-Yam, A., Maoz, D., Guhathakurta, P., \& Filippenko, A.~V.\ 2008, \apj, 680, 550 
\bibitem[gal-yam]{C}Gal-Yam, A., \& Leonard D.~C.\ 2009, \nat, 458, 865
\bibitem[Gilbank et al.(2008)]{2008ApJ...677L..89G} Gilbank, D.~G., Yee, 
H.~K.~C., Ellingson, E., Hicks, A.~K., Gladders, M.~D., Barrientos, L.~F., 
\& Keeney, B.\ 2008, \apjl, 677, L89 
\bibitem[Giodini et al.(2009)]{2009ApJ...703..982G} Giodini, S., et al.\ 
2009, \apj, 703, 982 
\bibitem[Girardi et al.(2000)]{2000AAS..141..371G} Girardi, L., Bressan, 
A., Bertelli, G., \& Chiosi, C.\ 2000, \aaps, 141, 371
\bibitem[Gonzalez et al.(2007)]{2007ApJ...666..147G} Gonzalez, A.~H., 
Zaritsky, D., \& Zabludoff, A.~I.\ 2007, \apj, 666, 147 
\bibitem[Gould et al.(1997)]{1997ApJ...482..913G} Gould, A., Bahcall, 
J.~N., \& Flynn, C.\ 1997, \apj, 482, 913 
\bibitem[Graham et al.(2008)]{2008AJ....135.1343G} Graham, M.~L., et al.\ 
2008, \aj, 135, 1343 
\bibitem[Greggio(2005)]		{2005A&A...441.1055G} Greggio, L.\
  2005, \aap, 441, 1055
\bibitem[Greggio et al.(2008)]{2008MNRAS.388..829G} Greggio, L., Renzini, 
A., \& Daddi, E.\ 2008, \mnras, 388, 829 
\bibitem[Greggio (2010)]{2010MNRAS.406..22G} Greggio, L. 2010, 
\mnras, 406, 22 
\bibitem[Grevesse \& Sauval(1999)]{1999A&A...347..348G} Grevesse, N., \& Sauval, A.~J.\ 1999, \aap, 347, 348 
\bibitem[Haines et al.(2009)]{2009ApJ...704..126H} Haines, C.~P., et al.\
2009, \apj, 704, 126
\bibitem[Han 
\& Podsiadlowski(2004)]{2004MNRAS.350.1301H} Han, Z., \& Podsiadlowski, P.\ 2004, \mnras, 350, 1301 
\bibitem[Hansen et al.(2009)]{2009ApJ...699.1333H} Hansen, S.~M., Sheldon, 
E.~S., Wechsler, R.~H., \& Koester, B.~P.\ 2009, \apj, 699, 1333 
\bibitem[Helder et al. (2009)]{2009Sci...325..719H} Helder, E.~A, et
  al. 2009, Sceince, 325, 719
\bibitem[Hicks et al.(2008)]{2008ApJ...680.1022H} Hicks, A.~K., et al.\ 
2008, \apj, 680, 1022 
\bibitem[Hoyle \& Fowler (1960)]{1960ApJ...132..565H} Hoyle, F., \&
  Fowler, W.~A.\ 1960, \apj, 132, 565
\bibitem[Iben 
\& Tutukov(1984)]{1984ApJS...54..335I} Iben, I., Jr., \& Tutukov, A.~V.\ 1984, \apjs, 54, 335 
\bibitem[Jorgensen et al.(1997)]{1997ApJ...486..110J} Jorgensen, H.~E., 
Lipunov, V.~M., Panchenko, I.~E., Postnov, K.~A., \& Prokhorov, M.~E.\ 
1997, \apj, 486, 110 
\bibitem[Kauffmann et al.(2003)]{2003MNRAS.341...33K} Kauffmann, G., et 
al.\ 2003, \mnras, 341, 33 
\bibitem[Koyama et al.(2010)]{2010MNRAS.403.1611K} Koyama, Y., Kodama, T.,
Shimasaku, K., Hayashi, M., Okamura, S., Tanaka, I.,
\& Tokoku, C.\ 2010, \mnras, 403, 1611
\bibitem[Krick et al.(2009)]{2009ApJ...700..123K} Krick, J.~E., Surace,
J.~A., Thompson, D., Ashby, M.~L.~N., Hora, J.~L., Gorjian, V.,
\& Yan, L.\ 2009, \apj, 700, 123
\bibitem[Kroupa et al.(1993)]{1993MNRAS.262..545K} Kroupa, P., Tout, C.~A., 
\& Gilmore, G.\ 1993, \mnras, 262, 545 
\bibitem[Kroupa(2001)]{2001MNRAS.322..231K} Kroupa, P.\ 2001, \mnras, 322, 
231 
\bibitem[Kuznetsova et al.(2008)]{2008ApJ...673..981K} Kuznetsova, N.,
  et al.\ 2008, \apj, 673, 981 
\bibitem[Lagan{\'a} et 
al.(2008)]{2008A&A...485..633L} Lagan{\'a}, T.~F., Lima Neto, G.~B., Andrade-Santos, F., \& Cypriano, E.~S.\ 2008, \aap, 485, 633 
\bibitem[leaman]{D} Leaman, J., Li, W., Chornock, R., Filippenko, A. 2010, 
ApJ, submitted, arXiv:1006.4611
\bibitem[li]{E1} Li, W., et al. 2010a, ApJ, submitted, arXiv:1006.4612 
\bibitem[li]{E2} Li, W., Chornock, R., Leaman, J., Filippenko, A.~V.,
Poznanski, D., Xiaofeng, W., Ganeshalingam, M., Mannucci, F. 2010b,
ApJ, submitted, arXiv:1006.4613 
\bibitem[Lin et al.(2003)]{2003ApJ...591..749L} Lin, Y.-T., Mohr, J.~J., 
\& Stanford, S.~A.\ 2003, \apj, 591, 749 
\bibitem[Loh et al.(2008)]{2008ApJ...680..214L} Loh, Y.-S., Ellingson, E., 
Yee, H.~K.~C., Gilbank, D.~G., Gladders, M.~D., 
\& Barrientos, L.~F.\ 2008, \apj, 680, 214 
\bibitem[Madau et al.(1998)]	{1998MNRAS.297L..17M} Madau, P., Della Valle, M., \& Panagia, N.\ 1998, \mnras, 297, L17 
\bibitem[Mannucci et al.(2005)]	{2005A&A...433..807M} Mannucci, F., Della Valle, M., Panagia, N., Cappellaro, E., Cresci, G., Maiolino, R., Petrosian, A., \& Turatto, M.\ 2005, \aap, 433, 807
\bibitem[Mannucci et al.(2006)]{2006MNRAS.370..773M} Mannucci, F., Della 
Valle, M., \& Panagia, N.\ 2006, \mnras, 370, 773 
\bibitem[Mannucci et al.(2008)]{2008MNRAS.383.1121M} Mannucci, F.,
  Maoz, D., Sharon, K., Botticella, M.~T., Della Valle, M., Gal-Yam,
  A., \& Panagia, N.\ 2008, \mnras, 383, 1121
\bibitem[Maoz(2008)]{2008MNRAS.384..267M} Maoz, D.\ 2008, \mnras, 384, 267 
\bibitem[Maoz et al.(2010)]{2010arXiv1002.3056M} Maoz, D., Mannucci, F., 
Li, W., Filippenko, A.~V., Della Valle, M., 
\& Panagia, N.\ 2010, arXiv:1002.3056, MNRAS, in press 
\bibitem[Maoz 
\& Badenes(2010)]{2010arXiv1003.3031M} Maoz, D., \& Badenes, C.\ 2010,
  arXiv:1003.3031, MNRAS, in press 
\bibitem[Maoz \& Gal-Yam(2004)]	{2004MNRAS.347..951M} Maoz, D., \& Gal-Yam, A.\ 2004, \mnras, 347, 951
\bibitem[Maughan et al.(2008)]{2008ApJS..174..117M} Maughan, B.~J., Jones, C., Forman, W., \& Van Speybroeck, L.\ 2008, \apjs, 174, 117
\bibitem[Mazzali et al.(2007)]{2007Sci...315..825M} Mazzali, P.~A., 
R{\"o}pke, F.~K., Benetti, S., \& Hillebrandt, W.\ 2007, Science, 315, 825 
\bibitem[Mennekens et al.(2010)]{2010arXiv1003.2491M} Mennekens, N., 
Vanbeveren, D., De Greve, J.~P., \& De Donder, E.\ 2010,
arXiv:1003.2491,\aap, in press 
\bibitem[Nelemans \& Tout(2005)]{2005MNRAS.356..753N} Nelemans, G., \& 
Tout, C.~A.\ 2005, \mnras, 356, 753 
\bibitem[Porter et al.(2008)]{2008MNRAS.388.1152P} Porter, S.~C., 
Raychaudhury, S., Pimbblet, K.~A., 
\& Drinkwater, M.~J.\ 2008, \mnras, 388, 1152 
\bibitem[Poznanski et al.(2007)]{2007MNRAS.382.1169P} Poznanski, D.,
  et al.\ 2007, \mnras, 382, 1169
\bibitem[Pratt et 
al.(2009)]{2009A&A...498..361P} Pratt, G.~W., Croston, J.~H., Arnaud,
  M., Boehringer, H.\ 2009, \aap, 498, 361 
\bibitem[Pritchet et al.(2008)]{2008ApJ...683L..25P} Pritchet, C.~J., 
Howell, D.~A., \& Sullivan, M.\ 2008, \apjl, 683, L25 
\bibitem[Raskin et al.(2009)]{2009ApJ...707...74R} Raskin, C., Scannapieco, 
E., Rhoads, J., \& Della Valle, M.\ 2009, \apj, 707, 74 
\bibitem[Reiss (2000)]{F} Reiss 2000, PhD thesis, University of Washington
\bibitem[Renzini(1997)]{1997ApJ...488...35R} Renzini, A.\ 1997, \apj, 488, 35
\bibitem[Renzini et al.(1993)]{1993ApJ...419...52R} Renzini, A., Ciotti, L., D'Ercole, A., \& Pellegrini, S.\ 1993, \apj, 419, 52
\bibitem[Rosati et al.(1999)]{1999AJ....118...76R} Rosati, P., Stanford, 
S.~A., Eisenhardt, P.~R., Elston, R., Spinrad, H., Stern, D., 
\& Dey, A.\ 1999, \aj, 118, 76 
\bibitem[Rosati et al.(2004)]{2004AJ....127..230R} Rosati, P., et al.\ 
2004, \aj, 127, 230
 \bibitem[Ruiter et al.(2009)]{2009ApJ...699.2026R} Ruiter, A.~J., 
Belczynski, K., \& Fryer, C.\ 2009, \apj, 699, 2026 
\bibitem[Saintonge et al.(2008)]{2008ApJ...685L.113S} Saintonge, A., Tran, 
K.-V.~H., \& Holden, B.~P.\ 2008, \apjl, 685, L113 
\bibitem[Sadat et 
al.(1998)]{1998A&A...331L..69S} Sadat, R., Blanchard, A., Guiderdoni, B., \& Silk, J.\ 1998, \aap, 331, L69 
\bibitem[Salpeter(1955)]{1955ApJ...121..161S} Salpeter, E.~E.\ 1955, \apj, 121, 161 
 \bibitem[Scannapieco \& Bildsten(2005)]{2005ApJ...629L..85S} Scannapieco, E., \& Bildsten, L.\ 2005, \apjl, 629, L85
\bibitem[Sharon et al.(2007)]{2007ApJ...660.1165S} Sharon, K., Gal-Yam, A., Maoz, D., Filippenko, A.~V., \& Guhathakurta, P.\ 2007, \apj, 660, 1165 
\bibitem[sharon]{G} Sharon, K., et al. 2010, ApJ, 718, 876 
\bibitem[Sim et al.(2010)]{jj}Sim, S.~A., et al. 2010, \apjl, 714, L52
\bibitem[Stanford et al.(2002)]{2002AJ....123..619S} Stanford, S.~A., 
Holden, B., Rosati, P., Eisenhardt, P.~R., Stern, D., Squires, G., 
\& Spinrad, H.\ 2002, \aj, 123, 619 
\bibitem[Stanford et al.(2005)]{2005ApJ...634L.129S} Stanford, S.~A., et 
al.\ 2005, \apjl, 634, L129 
\bibitem[Stanford et al.(2006)]{2006ApJ...646L..13S} Stanford, S.~A., et 
al.\ 2006, \apjl, 646, L13 
\bibitem[Strolger et al.(2004)]{2004ApJ...613..200S} Strolger, L.-G., et 
al.\ 2004, \apj, 613, 200 
\bibitem[Sullivan et al.(2006b)]{2006ApJ...648..868S} Sullivan, M., et
  al. 2006, \apj, 648, 868
\bibitem[Tojeiro et al.(2009)]{2009ApJS..185....1T} Tojeiro, R., Wilkins, 
S., Heavens, A.~F., Panter, B., \& Jimenez, R.\ 2009, \apjs, 185, 1 
\bibitem[Totani et al.(2008)]{2008PASJ...60.1327T} Totani, T., Morokuma, 
T., Oda, T., Doi, M., \& Yasuda, N.\ 2008, \pasj, 60, 1327 
\bibitem[Z]{z}van Kerkwijk, M.H., Chang, P., Justham, S. 2010,
 ApJL, submitted, arXiv:1006.4391
\bibitem[Wang et al.(2010)]{2010MNRAS.401.2729W} Wang, B., Li, X.-D., 
\& Han, Z.-W.\ 2010, \mnras, 401, 2729 
\bibitem[Webbink(1984)]{1984ApJ...277..355W} Webbink, R.~F.\ 1984, \apj, 
277, 355 
\bibitem[Whelan \& Iben(1973)]{1973ApJ...186.1007W} Whelan, J., \& Iben, 
I.~J.\ 1973, \apj, 186, 1007 
\bibitem[Woosley (2007)]{2007NatPh...3..832W} Woosley, S.~E.\ 2007,
  Nature Physics, 3, 832
\bibitem[Yasuda 
\& Fukugita(2010)]{2010AJ....139...39Y} Yasuda, N., \& Fukugita, M.\
  2010, \aj, 139, 39 
\bibitem[York et al.(2000)]{2000AJ....120.1579Y} York, D.~G., et al.\ 2000, 
\aj, 120, 1579
\bibitem[Yungelson \& Livio(2000)]{2000ApJ...528..108Y} Yungelson, L.~R., \& Livio, M.\ 2000, \apj, 528, 108
\end{thebibliography}
\end{document}